\documentclass[fleqn,usenatbib]{mnras}

\usepackage{newtxtext,newtxmath}

\usepackage[T1]{fontenc}

\DeclareRobustCommand{\VAN}[3]{#2}
\let\VANthebibliography\thebibliography
\def\thebibliography{\DeclareRobustCommand{\VAN}[3]{##3}\VANthebibliography}

\usepackage{graphicx}	
\usepackage{amsmath}	

\newcommand{\R}{R_{\lambda,q}}

\newcommand{\ben}{\begin{eqnarray*}}
\newcommand{\een}{\end{eqnarray*}}
\newcommand{\be}{\begin{eqnarray}}
\newcommand{\ee}{\end{eqnarray}}

\newcommand{\bit}{\begin{itemize}\setlength\itemsep{0em}}
\newcommand{\eit}{\end{itemize}}

\newcommand{\e}{\textit{e}}
\newcommand{\FD}{frame dragging\,}
\newcommand{\WK}{Wex-Kopeikin\,}
\newcommand{\RL}{Rafikov-Lai\,}

\newcommand{\SgrA}{SgrA$^{*}$}

\newcommand{\comment}[1]{}

\title[Rel. propagation and frame dragging time delay of a pulsar orbiting \SgrA]{Relativistic propagation and frame dragging time delay in the timing of a pulsar orbiting the supermassive black hole \SgrA}

\author[B.\ Ben-Salem \& E. Hackmann]{
Bilel Ben-Salem,$^{1,2}$\thanks{E-mail: bilelbensalem@physik.uni-bielefeld.de}
Eva Hackmann$^{2}$
\\
$^1$Fakul\"at f\"ur Physik, Universität Bielefeld, Postfach 100131, 33501 Bielefeld, Germany\\
$^2$ZARM, Universität Bremen, Am Fallturm 2, 28359 Bremen, Germany\\
}

\date{Accepted XXX. Received YYY; in original form ZZZ}

\pubyear{2022}

\begin{document}
\label{firstpage}
\pagerange{\pageref{firstpage}--\pageref{lastpage}}
\maketitle

\begin{abstract}
Timing a pulsar in a close orbit around the supermassive black hole \SgrA\ at the center of the Milky Way would open the window for an accurate determination of the black hole parameters and for new tests of General Relativity and alternative modified gravity theories. An important relativistic effect which has to be taken into account in the timing model is the propagation delay of the pulses in the gravitational field of the black hole. Due to the extreme mass ratio of the pulsar and the supermassive back hole we use the test particle limit to derive an exact analytical formula for the propagation delay in a Kerr spacetime and deduce a relativistic formula for the \FD\ effect on the arrival time. As an illustration, we treat an edge-on orbit in which the \FD\ effect is expected to be maximal. We compare our formula for the propagation time delay with  Post-Newtonian approaches, and in particular with the frame dragging terms derived in previous works by Wex \& Kopeikin and Rafikov \& Lai. Our approach correctly identifies the asymmetry of the frame dragging delay with respect to superior conjunction, avoids singularities in the time delay, and indicates that in the Post-Newtonian approach frame dragging effects are generally slightly overestimated. 
\end{abstract}


\section{Introduction}
Optical observations of star motions in the central region of the Milky revealed closed orbits around a supermassive black hole  with a mass of about $4.3\times 10^{6}$ M$_{\odot}$ \cite{Ghez2008} , \cite{Gillessen2009}, known as Sagittarius A$^{*}$ (\SgrA). 
This supermassive black hole offers an ideal laboratory  to test important features of black holes like the existence and shape of a back hole shadow \cite{Lu2014}, \cite{Falcke2000}, \cite{EHT2019i}, \cite{EHT2019ii}, \cite{EHT2019iii}, \cite{EHT2019vi}, the cosmic censorship conjecture \cite{Liu2012}, or the no-hair theorem \cite{Liu2012}, \cite{Psaltis2016}, \cite{Broderick2014}. It also offers the possibility  to test General Relativity and  modified gravity theories in a such strong gravity regime  never reached  before \cite{Liu2014},  \cite{Goddi2017}, \cite{Hees2017}, \cite{Mizuno2018} as well as alternative black hole candidates like boson stars \cite{Grould2017}, \cite{Cunha2017}, \cite{Vincent2016b}.

Similar to monitoring the $S2$ star with its orbital period of roughly 16 years \cite{Schoedel2002}, that has confirmed general relativity's prediction of Schwarzschild precession \cite{Gravity2020}, a possible discovery  of a pulsar closely orbiting the supermassive black hole would offer a promising laboratory  to explore the gravitational field of  \SgrA . A large population of millisecond pulsars at the order of 100  within the central parsec  \cite{Wharton2012}, \cite{Rajwade2017} is expected in the Galactic centre, whereas a higher pulsar number  based on recent observations is estimated in globular clusters \cite{Kramer2012}. A recent effort to search for pulsars in the Galactic center at higher frequency, at 2 and $3\, mm$, in order to reduce the scattering effect caused by the interstellar medium, failed to detect additional pulsars \cite{Torne2021}.

Even tough, the expected high sensitivity of the next generation radio telescopes such as the Square Kilometer Array \cite{Keane2014}, the Five-hundred-meter Aperture Spherical Telescope \cite{Nan2011}, or the NASA Deep Space Network \cite{Pearlman2019} have the potential to overcome  possible screening effects due to a high plasma density as well as strong magnetic fields in the galactic center.

Timing of a such pulsar orbiting \SgrA\ would offer an efficient approach to determine the physical and orbital parameters of the neutron star \cite{Verbiest2008}, \cite{Kramer2012}. A pulsar closely orbiting \SgrA would also enable to improve the accuracy of its mass estimate,  its spin magnitude and its corresponding orientation \cite{Liu2012}, \cite{Zhang2017}. Furthermore, it offers the possibility for the first time to test the no-hair theorem as well as  the cosmic censorship conjecture \cite{Liu2012}, \cite{Christian2015}, \cite{Izmailov2019}

In the usual binary systems the pulsar and the companion  have almost the same mass. However in the considered system the mass ratio is extreme what allows a pulsar orbit very close to the supermassive black hole and with a period of only a few years. In this case, relativistic effects are expected to be very strong not only on the orbit but also on the electromagnetic radiation \cite{Wang2008}, \cite{Wang2009}. The timing model that predicts the time of arrivals (ToAs) of the pulsar's radio waves is based on  Damour and Deruelle's approach using a post-Newtonian expansion to treat the relativistic two body problem \cite{Damour1986}, and its corresponding relativistic effects are described by a set of post-Keplerian parameters, see e.g. \cite{Edwards2006}, \cite{Damour1992}. The validity of the  post-Newtonian approximation that assumes a weak field becomes questionable for a pulsar orbiting closely a supermassive black hole, in particular if the black hole will be (nearly) in the line of sight to the pulsar .

In the treated binary system of a pulsar orbiting a supermassive black hole, the extreme mass ratio allows to consider another well motivated approximation, complementary to the post-Newtonian approach, namely the test particle limit. In this approximation the gravitational field of the pulsar is neglected, as compared to that of the black hole, and we may treat the motion of the pulsar and its radiation as geodesics in the gravitational field of the spinning black hole. In this approach, \cite{Hackmann2019} investigated the Shapiro delay as the dominant nontrivial relativistic effect on the radio pulses within the setting of a Schwarzschild spacetime that represents the non-spinning black hole case. In the present work, we generalize this previous treatment to include in addition to the relativistic Shapiro delay also the delay induced by the rotation, the \FD\ time delay, for the case of a black hole characterized by an additional parameter, the spin, within the frame of the Kerr spacetime. Of course it is in general not possible to disentangle the Shapiro delay and frame dragging terms in the form of a sum of two effects due to the nonlinearity of the relativistic equations; this is only possible in the weak field post-Newtonian approximations. We denote here as frame dragging delay the difference between the delay in a Kerr and a Schwarzschild spacetime, which represents not only the relativistic rotation delay but also includes all rotation induced multipole moments deviation (beginning by quadrupole moment) from a spherically symmetric configuration of the black hole \cite{Klioner1991}, \cite{Kopeikin1997}.

The outline of the paper is as follow. In section two we review the equations of motion for lightlike geodesics in a Kerr spacetime and solve for the propagation time from a given pulsar position  to an observer at infinity. We remark here that the employed solution method can be used to derive the time delay to an observer located at a finite position as well. Since in pulsar timing only the differences in the delay along the pulsar trajectory can be measured, and to enable to later compare the result in this approach with post-Newtonian expressions, we decided to fix the observer at infinity. In section three we find the finite exact propagation delay in Kerr spacetime with respect to a reference point. The weak field approximations of the propagation delay are reviewed in section four, up to the second post-Newtonian order. In the fifth section, we derived the relativistic \FD\ delay and review two  post-Newtonian based \FD expressions of Wex-Kopeikin and Rafikov-Lai. We represent in the following sixth section possible effects that could screen the direct measurement of \FD \ delay from pulsar timing and estimate their possible magnitudes. In the last seventh section,  we compare and discuss our results with the PPN approaches by taking as an example the most promising configuration, the edge-on case, where the pulsar and the observer are located in the equatorial plane, representing the galactic plane in the Milky Way. In this set up, the propagation as well as the \FD\ delays are expected to be maximal and therefore  it offers an occasion to test the accuracy of our derived formula. Finally, we close the paper with a summary and an outlook.

\section{Null geodesics in Kerr spacetime}
We use natural units such that $G = 1$, $c = 1$.
The line element of Kerr geometry in Boyer-Lindquist coordinates ($r$,$\vartheta$,$\varphi$,$t$) is given by \cite{Bardeen1972}
\be 
ds^2 = -e^{2\nu} dt^2 + e^{2\psi}(d\varphi - \omega dt)^2 + e^{2\mu_{1}} dr^{2} + e^{2\mu_{2}} d\vartheta^{2}\,,
\ee 
where the metric coefficients are given by
\begin{eqnarray*}
e^{2\nu}&=&\Delta \rho^2 A^{-1}\,,\\ 
e^{2\psi}&=&\sin^2{\vartheta} A \rho^{-2}\,,\\
e^{2\mu_{1}}&=& \rho^{2} \Delta^{-1}\,,\\
e^{2\mu_{2}}&=&\rho^2\,,\\
\omega&=&2 M a r A^{-1}\,,
\end{eqnarray*}
with the following definitions
\ben 
\Delta &=& r^2 -2 M r +a^2\,,\\
\rho^2 &=& r^2 + a^2 \cos^2\vartheta\,, \\
A &=& (r^2 + a^2)^2 - a^2\Delta \sin^2\vartheta\,.
\een
Here the parameter $M=Gm/c^2$ is related to the mass $m$ of the black hole, and $a=J/(Mc)$ is related to the angular momentum $J>0$.

Photons move along null geodesics which satisfy the geodesic equation
\be 
0 = \ddot{x}^{\mu} + \Gamma^{\mu}_{\nu\rho} \dot{x}^{\nu}\dot{x}^{\rho}
\ee 
where $\Gamma$ defines the Christoffel symbols and the dot denotes the derivative with respect to an affine parameter $\tau$ along the curve.  
From stationarity and axial symmetry we can infer two constants of motion,
\begin{align}
    E & = g_{00} u^0 + g_{0\varphi} u^\varphi\,,\\
    L & = g_{\varphi\varphi} u^\varphi + g_{\varphi 0}u^0,,
\end{align}
related to the energy and angular momentum of the particle. A third constant is given by the condition for photons $g_{\mu\nu} u^\mu u^\nu = 0$. It was \emph{Brandon Carter} who found the fourth constant of motion by working in the Hamiltonian formalism. The Carter constant $Q$ appears as a separation constant for the Hamilton-Jacobi equation.\cite{Carter1974}

 The equations of motion for photons following null geodesics in Kerr spacetime are \cite{O'Neill2014}
 \begin{eqnarray}
 \frac{dt}{d\gamma} &=& \frac{(r^2+a^2)(r^2+a^2-a\lambda)}{\Delta} -a(a-\lambda) - a^2\cos^2\vartheta  
 \,, \label{eqom1} \\
\frac{d\varphi}{d\gamma} &=& \frac{a(r^2+a^2-a\lambda)}{\Delta} - a +\frac{\lambda}{\sin^2\vartheta}
\,, \label{eqom2} \\
\left(\frac{d\vartheta}{d\gamma}\right)^2 &=& q + a^2\cos^2\vartheta - \lambda^2 \cot^2\vartheta =: \Theta_{\lambda,q}(\vartheta) \,, \label{eqom3}  \\
\left(\frac{dr}{d\gamma}\right)^2 &=& (r^2+a^2 - a \lambda)^2 - \Delta (q+(\lambda-a)^2) =: R_{\lambda,q}(r) \,. \label{eqom4}
\end{eqnarray} 
Here the auxiliary \emph{Mino parameter} $\gamma$ was introduced to completely decouple the equations. It is related to an affine parameter $\tau$ by $d\tau/d\gamma = \rho^2$. Moreover, we used the notation
\begin{equation}
q = \frac{Q}{E^2} \,, \quad \lambda = \frac{L}{E} \,.
\end{equation}  

Let us quickly review the possible types of motion that can appear for photon orbits in Kerr spacetime. The parameter $q$ determines the type of latitudinal motion: for $q>0$ the motion oscillates around the equatorial plane, for $q=0$ the orbit is confined to the equatorial plane, and for $q<0$ the photon is bounded away from the equatorial plane and oscillates between two latitudes $0<\vartheta_{\rm min} <\vartheta <\vartheta_{\rm max}<\pi/2$ in the northern hemisphere (with an equivalent symmetric orbit in the southern hemisphere). As the ring singularity is located in the equatorial plane, orbits with $q<0$ might possibly pass through the ring and can be continued to negative $r-$coordinates. We will denote such motions as crossover orbits. We are however only interested in the part of the orbit that is located outside of the two horizons.

The radial motion is governed by the quadratic polynomial $R_{\lambda,q}(r)$ that has four roots
\be\label{eq:R-roots} 
R_{\lambda,q}(r)= (r-r_{1})(r-r_{2})(r-r_{3})(r-r_{4}) 
\ee   
with $r_{i}<r_{i+1}$ if all roots are real. As $R_{\lambda,q}$ needs to be positive, depending on the number of real zeros we identify three different types of null orbits \cite{Hackmann2010} :
\begin{enumerate}[(I)] 
\item All zeros of $\R$ are complex, leading to transit orbits. This type of radial motion is only possible for $q<0$.   
\item $\R$ has two real zeros $r_{1}$, $r_{2}$ and $\R(r)\geq 0$ for $ r \leq r_{1}$ and $r_{2} \leq r$. Possible orbit types : two flyby orbits, one to $+\infty$ and one to $-\infty$. 
\item All four zeros of $\R$ are real and $\R(r)\geq 0$ for $r \leq r_{1}$, $r_{2} \leq r \leq r_{3}$ and $r_{4} \leq r$. Possible orbit types: two flyby orbits, one to each of $\pm \infty$ and a bound orbit.

\end{enumerate} 
The flyby orbit between $r=-\infty$ and $r_1<0$ in both (II) and (III) lies entirely in the negative radial coordinates and is therefore of no interest to us. The bound orbit in (III) is to be understood in the sense that it is confined to a finite radial region. It will however cross both horizons. A border case between (II) and (III) is given by $r_4=r_3$, and corresponds to the unstable spherical photon orbits. 

They are the only photon trajectories that neither cross the horizons nor escape to infinity (see e.g \cite{Chandrasekhar1998} ). Their constant radius $r_{c}$ is characterized by  
\be 
\lambda_{c}(r_{c}) &=& \frac{1}{a(r_{c} - M)}(M (r_{c}^2 -a^2) - r_{c} \Delta_{c}) \\
q_{c}(r_{c}) &=& \frac{r_{c}^3}{a^2(r_{c} - M)^2}(4M\Delta_{c} - r_{c}(r_{c} - M)^2)
\ee 
where the range of $r_{c}$ is limited by 
\ben 
r^{+}_{ph} \leq r_{c} \leq r^{-}_{ph} \leq 4M
\een
and 
\be 
r^{\pm}_{ph} = 2 M \left(1 + \cos\Biggl[\frac{2}{3}\cos^{-1}\Biggl(\frac{\pm a}{M}\Biggr)\Biggr] \right)
\ee 
corresponds to prograte and retrograde circular null orbits.

For the special case $q=0$ that corresponds to an orbit in the equatorial plane, we note that the roots $r_{i}$ of $\R$ are all real for the case that $\lambda > \lambda^{+}_{c}=\lambda_{c}(r^{+}_{ph})$ and $\lambda < \lambda^{-}_{c}=\lambda_{c}(r^{-}_{ph})$. Then $r_{1}$ is always negative, whereas the two  positive roots $r_{3}, r_{4}$ will merge to a double root for $\lambda = \lambda^{\pm}_{c}$ and become a complex pair for $\lambda^{-}_{c}<\lambda<\lambda^{+}_{c}$.

We are interested in the arrival times of photons $t_a$ as measured by an observer at infinity.  From ~(\ref{eqom1}) (\ref{eqom3}) and  (\ref{eqom4}) we get the equation for $t$ in integral form 
\begin{align}\label{eq:Int-t} 
(t_{a} - t_{e}) &= \int_{\gamma_{r}} \frac{r^2(r^2+a^2)+2Mar(a-\lambda)}{\Delta \sqrt{R_{\lambda,q}(r)}} dr \nonumber\\
&+ \int_{\gamma_{\vartheta}} \frac{a^2\cos^2\vartheta}{\sqrt{\Theta_{\lambda,q}(\vartheta)}} d\vartheta
\end{align} 
where $t_e$ is the coordinate time of emission. The integration path $\gamma_r$ starts at the radial point of emission $r_e$ and either runs directly to infinity or first decreases in radius until a turning point $r_4$ is reached from where it then proceeds to infinity.
The radial part of the integral in (\ref{eq:Int-t}) can be written as 
\be\label{eq:Int-r-2}
\Biggl(\int_{r_{4}} ^{\infty} \pm \int_{r_{4}} ^{r_{e}}\Biggr) \frac{r^2(r^2+a^2)+2Mar(a-\lambda)}{\Delta \sqrt{R_{\lambda,q}(r)}} dr\,.
\ee
For monotonically increasing $r$ we choose the minus sign in (\ref{eq:Int-r-2}) and else the plus sign. We note that for the case that not all zeros are real, $r_{4}$ could be one of the two complex conjugate roots but the complete expression is still real.

For the second integral, the integration path $\gamma_\vartheta$ begins at the angle of emission $\vartheta_e$ and runs to the arrival position at the observer $\vartheta_a$, possibly passing through one or more turning points. We can rewrite the latitudinal part by introducing a new variable $u=\cos^2(\vartheta)$ in the following form
\be\label{eq:int-u}
 \int_{\gamma_{\vartheta}} \frac{a^2\cos^2\vartheta}{\sqrt{\Theta_{\lambda,q}(\vartheta)}} d\vartheta & = &\int_{u_{e}} ^{u_{a}} \frac{a^2}{2} \frac{u du}{\sqrt{U_{\lambda,q}}}\,,\\
\text{where} \qquad U_{\lambda,q}(u) &=& u (q + u (a^2 - \lambda^2 - q) - a^2 u^2)\,.
\ee 

We remark that the sign of the square root has to be chosen according to the direction of motion. For the radial part, the positive sign corresponds to radial outwards motion. Concerning the latitudinal motion towards the south pole with increasing values of $\vartheta$, one must take the positive sign, motion towards the north pole requires the negative sign.\\

The integral (\ref{eq:Int-t}) can be solved exactly in terms of elliptic integrals. Details of the derivation can be found in appendix \ref{Appendix}. The result is  
\be
t_{a} - t_{e}=\frac{M}{c} \left[ T_r(\infty,\lambda_e,q_e) \pm T_r(r_{e},\lambda_e,q_e) + T_u(u_a,\lambda_e,q_e)\right] \label{eq:tate}
\ee
where
\be
&&T_r(r,\lambda,q) = \int_{r_4}^{r} \frac{r^2(r^2+a^2)+2Mar(a-\lambda)}{\Delta \sqrt{R_{\lambda,q}(r)}} dr \nonumber \\ 
&&=\frac{2M}{\sqrt{(r_{3}-r_{1})(r_{4}-r_{2})}} \Biggl[\Biggl(4M + \frac{r_{3}r_{4} - r_{1}r_{4}+r_{1}r_{3}+r_{3}^2}{2M} + 2r_{4} \nonumber \\
&&+ \frac{(a\lambda - 4 M^2)r_{-} - 2Ma^2}{(r_{3}-r_{-})\sqrt{M^2 - a^2}} -\frac{(a\lambda - 4 M^2)r_{+} + 2Ma^2}{(r_{3}-r_{+})\sqrt{M^2 - a^2}}\Biggr) F(x,k)\nonumber \\
&& - \frac{(r_{3}-r_{1})(r_{4}-r_{2})}{2M} E(x,k) - 2(r_{4}-r_{3}) \Pi(x,c_{3},k) \nonumber \\
&&+\frac{r_{4} - r_{3}}{\sqrt{M^2-a^2}}\Biggl(\frac{r_{+}a\lambda - 4r_{+}M^2 + 2Ma^2}{(r_{3} - r_{+})(r_{4} - r_{+})} \Pi(x,c_{+},k) \\ 
&&- \frac{r_{-}a\lambda - 4r_{-}M^2 + 2Ma^2}{(r_{3} - r_{-})(r_{4} - r_{-})}\Pi(x,c_{-},k)\Biggl)\Biggr] \nonumber \\
&&+ \frac{\sqrt{\R}}{r-r_{3}} + 2M ~\text{ln}\Biggl(\frac{\sqrt{(r-r_{2})(r-r_{1})}+\sqrt{(r-r_{4})(r-r_{3})}}{\sqrt{(r-r_{2})(r-r_{2})}-\sqrt{(r-r_{4})(r-r_{3})}}\Biggr)\nonumber\\
&&T_u(u,\lambda,q) = \int_{u_{e}}^{u} \frac{a^2}{2} \frac{u du}{\sqrt{U_{\lambda,q}}} \nonumber\\
&& = \frac{I~\text{sign}(a)}{\sqrt{u_{+} - u_{-}}} \Biggl( (u_{-} - u_{+}) E(v,w) + u_{+} F(v,w) \Biggr)   
\ee
The indices in $\lambda_e$, $q_e$ in eq.~\eqref{eq:tate} indicate here, that the constants of motion depend on the emission point $(r_e,\varphi_e,\vartheta_e)$. Generally they of course also depend on the observer position, which we fixed here to $r_a=\infty$ and $\varphi_a=0$ (whereas $u_a=\cos^2\vartheta_a$ remains free); however, completely analogously one can choose another observer position.

In the above equations, $r_i$ are the zeros of $R_{\lambda,q}$ as before, $r_{\pm}$ are the horizons, the constants $c_1$, $c_2$, $c_{\pm}$, and $k$ are related to $r_i,r_{\pm}$ and defined in equations \eqref{eq:defc}, \eqref{eq:defk}, and $x$ is a function of $r$ defined in eq.~\eqref{eq:defx}. The non-zeros roots of $U_{\lambda,q}$ are denoted by $u_{\pm}$, and $v$, $w$ are defined in eqs.~\eqref{eq:defv}, \eqref{eq:defw}.


\section{The propagation time delay in Kerr spacetime}

\subsection{The exact propagation time delay formula}

In order to calculate the time delay in \eqref{eq:Int-t} we need the coordinates of the emission point $(r_e,\vartheta_e,\varphi_e,t_e)$ on the pulsar orbit as well as the observer inclination $\vartheta_a$ (we assume $r_a=\infty$, $\varphi_a=0$). From that we need to determine the parameters of the photon geodesic $(q,\lambda)$ that connect the emission point with the observer. The problem of determining this connecting light ray is known as the emitter-observer problem. There is to our knowledge no general analytical solution. Note that in general there are of course infinitely many lightlike geodesics connecting emitter and observer; here we always choose the primary, i.e. the one with the shortest time delay. 

Therefore, in general we need to determine the parameters $(q,\lambda)$ of the connecting light ray numerically from the equations
\begin{align}
\int_{\gamma_{r}} \frac{dr}{\sqrt{R_{\lambda,q}(r)}}&=\int_{\gamma_\vartheta} \frac{d\vartheta}{\sqrt{\Theta_{\lambda,q}(\vartheta)}}\,, \label{eq:int-rtheta}\\
\varphi_{a}-\varphi_{e}&=\int_{\gamma_{r}} \frac{2Mar - \lambda a^{2}}{\Delta \sqrt{R_{\lambda,q}(r)}} dr + \int_{\gamma_{\vartheta}} \frac{\lambda}{\sin^2\vartheta \sqrt{\Theta_{\lambda,q}}}d\vartheta \label{eq:int-phi_e}
\end{align}
For a detailed discussion of the first equation we refer to \cite{Viergutz1993}. In his work he identified valid parameter regions of $\lambda$ and $q$ for given emitter and observer positions and computed the one-dimensional line in the two-dimensional ($\lambda,q)$ plane that corresponds to all possible $\varphi_e$. In addition to his work, here we then need to find the particular value of $(\lambda,q)$ on this line corresponding to the given $\varphi_e$ of the pulsar. 

However, in the particular case that everything is restricted to the equatorial plane the problem is much simpler, and at the same time most interesting, as this case corresponds to the strongest relativistic effects. In this case we can choose $q=0$ and do not need to care about \eqref{eq:int-rtheta} but only about the $\varphi$ coordinate. Eq.~\eqref{eq:int-phi_e} then simplifies to
\be
\varphi_{a} - \varphi_{e}&=&\int_{\gamma_{r}} \frac{2Mar - \lambda a^{2}}{\Delta \sqrt{R_{\lambda,q=0}(r)}} dr\nonumber\\
&=&\Biggl( \int_{r_4}^\infty \pm \int_{r_4}^{r_e} \Biggr) \frac{2Mar - \lambda a^{2}}{\Delta \sqrt{R_{\lambda,q=0}(r)}} dr 
\ee
From this equation we then need to find the remaining parameter $\lambda$ for given $\varphi_e$.

For circular equatorial orbits of the pulsar, we can actually give a grid of $\lambda$ values and it is straigthforward to determine the corresponding $\varphi$ coordinate. In more general setups, like eccentric and/or inclined orbits we numerically determined $\lambda$. Here it would be very helpful to develop accurate analytical approximation methods, as available for Schwarzschild spacetime (see \cite{Semerak2015} for a review), but this is beyond the scope of the this paper.

The relativistic propagation delay is given as the difference between the time delay of a signal from the actual position of the pulsar and a chosen reference point 
\begin{align}\label{eq:prop_delay}
&\Delta t_{\text{ex}}(r_{e},\varphi_{e},u_{e}) = (t_{a}-t_{e}) - (t_{a}-t_{\text{ref}})\nonumber \\
&= \frac{M}{c} \biggl[T_r(\infty,\lambda_{e},q_{e})\pm T_r(r_{e},\lambda_{e},q_{e}) + T_u(u_a,\lambda_e,q_e) \Biggr]  \\
&- \frac{M}{c}  \Bigg[ T_r(\infty,\lambda_{\text{ref}},q_{\text{ref}}) \pm T_r(\infty,\lambda_{\text{ref}},q_{\text{ref}}) + T_u(u_a,\lambda_{\text{ref}},q_{\text{ref}}) \Bigg]  \nonumber 
\end{align}
where the indices in $\lambda_{e,\text{ref}}$ and $q_{e,\text{ref}}$ indicate the dependence on the emission and reference position. Note that the diverging terms in $T_r(\infty,\lambda_{e},q_{e})$ and $T_r(\infty,\lambda_{\text{ref}},q_{\text{ref}})$ directly cancel each other (see appendix \ref{Appendix} ). Therefore we are left only with a finite difference between the arrival times of signals from the pulsar as it orbits the central rotating supermassive black hole. 

\begin{figure}
	\centering
	\includegraphics[width=0.4\textwidth]{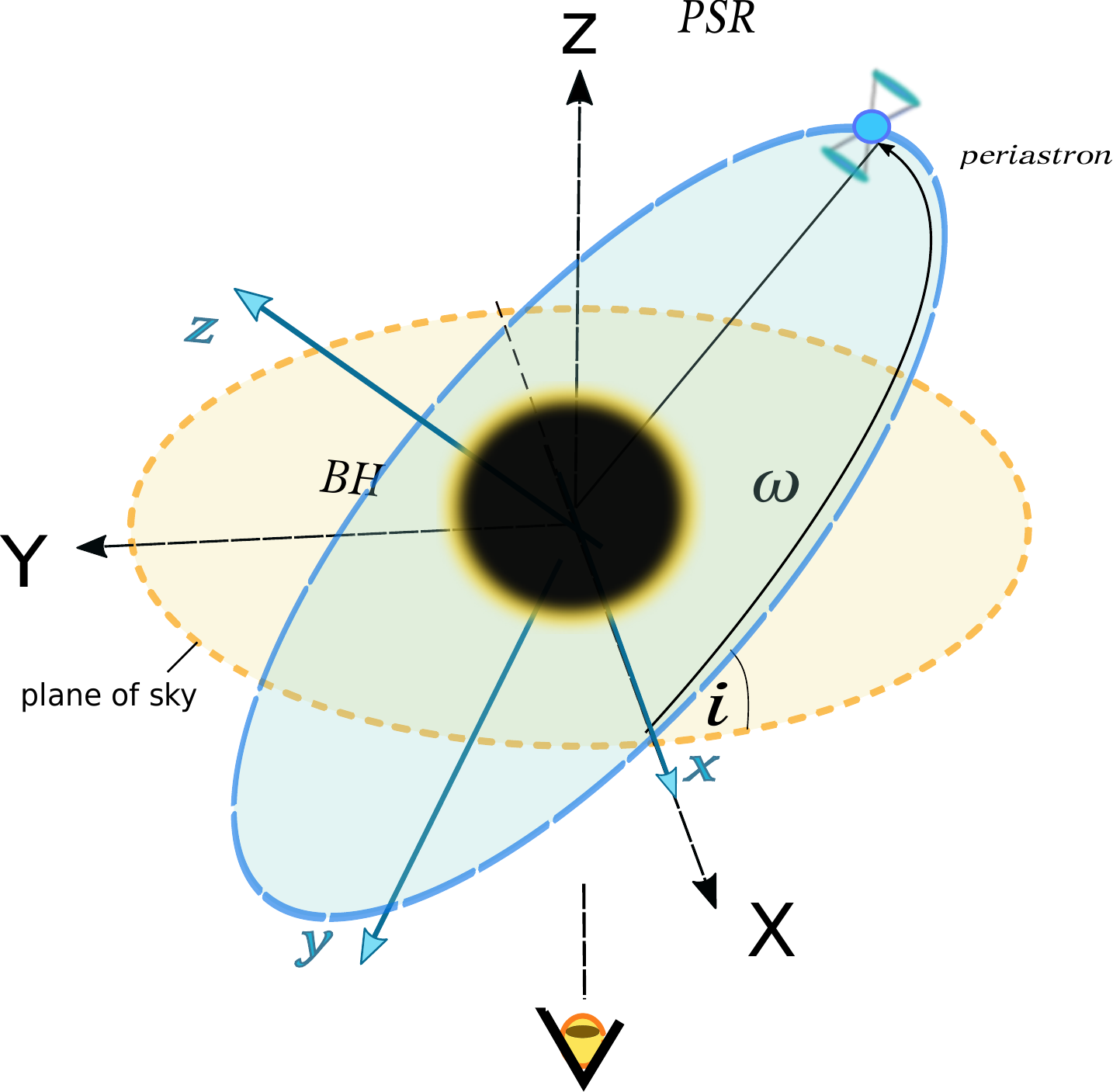}
	\caption{Orientation of the black hole-pulsar system in the sky with respect to an observer sitting at infinity.}
	\label{fig:diagram1}
\end{figure}

Let us now relate the coordinates used in the above equations to the coordinate systems usually used in Pulsar Timing models. To this end we review the  geometrical set-up already introduced in \cite{Hackmann2019} that we adopt in this paper. Due to the big difference in the masses of a pulsar orbiting a supermassive black  hole like \SgrA, we may consider the pulsar as a test particle and the center of mass to coincide with the center of the black hole. We set the origin of our coordinate system $(X,Y,Z)$ to the center of the black hole in such a way that the $Z$-axis is fixed by the line of sight from the observer to the origin. In this case the ascending node of the pulsar orbit with respect to the plane of sky  determines the $X$-axis. The inclination $\textit{i}$ of the pulsar orbit is measured with respect to the plane of sky that coincides with the $X$-$Y$-plane of our coordinate system (see figure \ref{fig:diagram1}).

We have to mention here that of course the pulsar will in general not remain in a fixed plane due to the frame dragging effects on its orbit, and also not on a fixed ellipse due to the relativistic precession of the periapsis. We nevertheless identify here a Keplerian orbit for the pulsar, as common in post-Newtonian Pulsar Timing expressions. Relativistic effects on the pulsar orbit will be encoded in the post-Keplerian orbital parameters, that are accounted for separately in the timing model. Our goal is here to relate the Boyer-Lindquist coordinates used in our formula to the coordinates used in the post-Newtonian expression in the next section. To this end, we assume here a  Keplerian orbit. Note that this does not have any impact on our solution method that is valid for any given emission position outside the event horizon. In a complete pulsar timing model, of course the post-Keplerian parameters for the pulsar orbit will generally induce time varying orbital elements.

We choose $(x,y,z)$ to be the coordinate system in which the pulsar is in the $x-y$-plane with $x=X$. The pulsar motion can then be described by $r_{e}= \frac{A(1-e^2)}{1+e\cos\phi}$ where $A$ is the semi-major axis, $e$ the eccentricity, and $\phi$ is the true anomaly. In this case we can express the coordinate $(x,y,z)$ as follows: $x= r_{e}\cos(\omega+\phi)$, $y=r_{e}\sin(\omega+\phi)$, $z=0$, where $\omega$ is the argument of the periastron. A rotation around the $x$-axis by the inclination angle $i$ suffices to transform to the $(X,Y,Z)$ system. For the case of an edge-on equatorial pulsar orbit, that we will discuss later in the paper, the desired angle between pulsar and observer is then given by the angle $\vartheta$ in spherical coordinates $X= r\cos\psi\sin\vartheta$, $Y=r\cos\psi\sin\vartheta$, $Z=r\cos\vartheta$. In the common plane of pulsar and observer, the angle $\phi_{e}$ is then determined by $\varphi_{e}=\vartheta$ with $\cos\vartheta= -\sin\textit{i}\sin(\omega+\phi)$ and therefore
\be\label{eq:phi} 
\cos\varphi_{e} = -\sin\textit{i}\sin(\omega+\phi)\,.
\ee


\subsection{Frame dragging time delay} \label{subsec:FD}

The relativistic frame dragging time delay can be defined as the difference between the exact propagation time delay in Kerr and Schwarzschild spacetimes as follows,
\be\label{eq:FD-Full} 
\Delta t_{\text{FD}} &= \Delta t_{\text{ex}} - \Delta t_{a=0}
\ee
where the delay in Schwarzschild spacetime was derived in \cite{Hackmann2019},

\begin{align}
&\Delta t_{a=0}=\frac{2}{(r_{4}'-r_{2}')(r_{3}'-r_{1}')} \Biggl[\Biggl(\frac{r_{3}'^3}{r_{3}'-2} + \frac{(r_{4}'-r_{3}')(r_{3}'-r_{1}')}{2} \Biggr)F(x',k')\nonumber\\
&-\frac{(r_{4}'-r_{2}')(r_{3}'-r_{1}')}{2}E(x',k') + 2 (r_{4}'-r_{3}') \Pi(x',c_{1}',k') \\
&- \frac{8M(r_{4}'-r_{3}')}{(r_{4}'-2)(r_{3}'-2)} \Pi(x',c_{2}',k')\Biggr] - \frac{\sqrt{R_{\lambda,q}'}}{(r-r_{3}')} \nonumber\\ 
&-2M \ln\left(\frac{\sqrt{(r-r_{2}')(r-r_{1}')}+\sqrt{(r-r_{4}')(r-r_{3}')}}{\sqrt{(r-r_{2}')(r-r_{1}')}-\sqrt{(r-r_{4}')(r-r_{3}')}}\right)\nonumber
\end{align}

Here 
\begin{align}
R'_{\lambda,q}(a=0)&=(r-r_{1}') (r-r_{2}') (r-r_{3}') (r-r_{4}'),\nonumber\\
k'^2&=\frac{(r_{3}'-r_{2}')(r_{4}'-r_{1}')}{(r_{4}'-r_{2}')(r_{3}'-r_{1}')}, \quad x'^{2} = \frac{(r-r_{4}')(r_{3}'-r_{1}')}{(r-r_{3}')(r_{4}'-r_{1}')},\nonumber\\
c_{1}'&=\frac{r_{4}'-r_{1}}{r_{3}'-r_{1}'} , \quad c_{2}'=\frac{(r_{4}'-r_{1}')(r_{3}'-2)}{(r_{3}'-r_{1}')(r_{4}'-2)}\nonumber
\end{align}

where $'$ denotes the radial coordinate in Schwarzschild spacetime. 

As the Boyer-Lindquist coordinates used in \eqref{eq:FD-Full} do not have an intrinsic physical meaning, we need to identify a physical invariant in order to compare results derived in different spacetimes. The circumference of a circle is such an invariant characteristic. In the Schwarzschild spacetime the circumference is given by $2\pi r_{\text{Sch}}$, where $r_{\text{Sch}}$ is the usual Boyer-Lindquist type radial coordinate of the Schwarzschild metric. In the Kerr spacetime, the circumference is $2\pi r_{\text{Kerr}}\sqrt{1+a^2/r^2_{\text{Kerr}}+2Ma^2/r^3_{\text{Kerr}}}$. We choose to keep this circumference fixed, which then implies that
\be \label{eq:rSK}
r_{\text{Schw}} = \frac{r_{\text{Kerr}}}{\sqrt{(1+a^2/r_{\text{Kerr}}^2+2 M  a^2/r_{\text{Kerr}}^3)}}.
\ee 
Note that we could as well have chosen another invariant characteristic of the pulsar orbit. However, already in \cite{Hackmann2019} it was pointed out that the circumference of a circle works well also in the comparison to post-Newtonian approaches, in contrast to other choices as, say, an identification of the proper orbital period.

We remark here that in the Schwarzschild case there is no $\vartheta$ contribution to the frame dragging delay $\Delta t_{\text{FD}}$ because of the spherical symmetry of the metric.

\section{The propagation time delay in the weak field } \label{sec:weakfield}

\subsection{Propagation time delays without frame dragging effects}
Here we review the post Newtonian approximation of the propagation time delay from literature.
\subsubsection{The Roemer and Shapiro time delay}
Based on the work of Blandford and Teukolsky \cite{Blandford1976} the Roemer delay is given by :
\be\label{eq:Roemer} 
\Delta t_{\text{R}} = \frac{a(1-\textit{e}^2)\sin\textit{i}\sin(\omega+\phi)}{c(1+\textit{e}\cos\phi)}
\ee 
where $\textit{i}$ is the inclination of the orbital plane with respect to the plane of sky and $\omega$ is the argument of periapsis, $\phi$ is the argument of the pulsar's position and $\textit{e}$ is the eccentricity of the orbit. The Shapiro delay is :
\be\label{eq:shapiro}
\Delta t_{\text{S}}= \frac{2GM}{c^3} \ln \Biggl( 
\frac{1+\textit{e}\cos\phi}{1-\sin\textit{i}\sin(\omega + \phi)}\Biggr)
\ee  
The Roemer delay vanishes at $\phi = -\omega$, whereas the Shapiro delay vanishes at 
\be 
\phi = \arctan\Biggl(\frac{-\textit{e}-\sin\textit{i}\sin\omega}{\sin\textit{i}\cos\omega}\Biggr)
\ee 
For circular orbits with $\textit{e} = 0$ we also find $\phi = -\omega$, i.e the ascending node. The point where the time delay vanishes can be considered as the reference point. The formula (\ref{eq:shapiro}) diverges for edge-on trajectories with $\textit{i} = \frac{\pi}{2}$ at superior conjunction, $\omega + \phi = \frac{\pi}{2}$. This divergence can be explained that a straight path of light that passes through a central object, where we have an infinitely deep gravitational potential. To circumvent this, we take the lensing path into account that was  first  introduced by Schneider \cite{Schneider1990} and later corrected by Rai and Rafikov \cite{Lai2005}. The generalized result is:
\be\label{eq:shapiro-lensed}
\Delta t_{\text{S,l}} = \frac{2GM}{c^3} \ln \Biggl( \frac{a(1-e^2)}{\sqrt{|\Vec{r}_{e}\cdot\Vec{n}|^2}+|\Vec{r}_{\pm}|^2 - \Vec{r}_{e}\cdot\Vec{n}} \Biggr)
\ee 
where $\Vec{r}_{\pm}$ is the approximate position of the image of the source in the plane of sky, 
\be 
\Vec{r}_{s}=\Vec{r}_{e}\sqrt{1-\sin^2\textit{i}\sin^2(\omega_\phi)}
\ee 
Here $R_{E}$ denotes the Einstein radius, which can be approximated by $R_{E}^2 = \frac{4GM}{c^2}|\Vec{r}_{e}|\sin\textit{i}$ at superior conjunction $|\Vec{r}_{e}| = \frac{a(1 - e^2)}{1+e\sin\omega}$.
\subsubsection{The geometric delay}
The geometric delay is the additional travel time that the light ray needs along a curved path in  the gravitational potential of the black hole compared to a straight path. This delay becomes significant when the pulsar is on the farther side of the orbit around the black hole. To first order the path can be approximated as a straight line from the point of emission to the minimum distance to the black hole and then from there to the observer. In this case, the delay in the difference between this path length and the straight line path from pulsar to the observer.\\
To first order the geometric delay is given by \cite{Lai2005}
\be 
\Delta t_{\text{geo}} = \frac{2GM}{c^3} \Biggl( \frac{|\Vec{r}_{\pm} - \Vec{r}_{s}|}{R_{E}} \Biggr)^2
\ee 
If $R_{E}$ is large compared to $|\Vec{r}_{s}|$ we have $|\Vec{r}_{\pm} - \Vec{r}_{s}| \rightarrow R_{E}$ and the two images merge into an Einstein ring. In the other case where $|\Vec{r}_{s}|>> R_{E}$ we find $|\Vec{r}_{+}-\Vec{r}_{s}|\rightarrow R_{E}^2/|\Vec{r}_{s}|$ and $|\Vec{r}_{-}-\Vec{r}_{s}|\rightarrow |\Vec{r}_{s}|$, but the "-" image is very faint therefore negligible.
\subsubsection{The second order Shapiro delay}
Based on the work of Zschocke and Klioner \cite{Zschocke2009},the second post-Newtonian order of the Shapiro delay can be expressed as 
\be 
\Delta t_{2\text{PN}} = \frac{GM}{c^3 r} \Biggl(-\frac{4}{1+\cos\varphi_{e}} + \frac{\cos\varphi_{e}}{4} + \frac{15\varphi_{e}}{4\sin\varphi_{e}}\Biggr)
\ee 
where $\varphi_{e}$ is the angle between the emission position vector $\Vec{r}_{e}$ and receiver position vector $\Vec{r}_{a}$ and $r = a(1-e^2)/(1+e\cos\phi)$. From the formula we notice that at inferior conjuction $\varphi_{e} = 0$ the last term is finite, whereas at superior conjunction $\varphi_{e} = \pi$ the first term diverges.
\label{sec:FD}

\subsection{Frame dragging time delays}
\subsubsection{Wex-Kopeikin time delay}
In \cite{Wex1999} Wex and Kopeikin derived an analytical approximate expression for the frame dragging time delay based on perturbative methode for a photon trajectory near the black hole and where the observer is considered to be at a large distance. This approach can be regarded as a special case ($l=1$) of a general treatment in \cite{Kopeikin1997}, in the context of gravitational lenses, for the propagation of light rays in the stationary field of relativistic gravitational multipoles. 
Using the angles in figure \ref{fig:diagram1}, the FD propagation effect in a binary-pulsar system is given by:
\be\label{eq:W-K} 
\Delta t_{\text{FD}} = \Lambda^{-1} \Biggl[A_{\text{FD}}\cos\textit{i}\sin(\omega + A_{e}(u)) + B_{FD} \cos(\omega + A_{e}(u))\Biggr]
\ee 
where the function $\Lambda$ is defined by
\be 
\Lambda = 1 - \textit{e}\cos u - s [\sin\omega(\cos u - \textit{e}) + (1 - \textit{e}^2)^{1/2}\cos\omega\sin u] \,.
\ee 
The eccentric anomaly angle is 
\be 
A_{\e}(u) = 2 \arctan\Biggl[ \Biggl(\frac{1+\e}{1-\e}\Biggr)^{1/2} \tan\frac{u}{2}\Biggr]
\ee 
and 
\be 
\textit{A}_{FD} &=& +4 T_{\odot}^{5/3} \Biggl(\frac{2\pi}{P_{b}}\Biggr) \frac{M_{\bullet}^2}{(M_{p} + M_{\bullet})^{1/3}} \chi \sin\lambda_{\bullet}\cos\nu_{\bullet} \,,\\
\textit{B}_{FD} &=& -4 T_{\odot}^{5/3} \Biggl(\frac{2\pi}{P_{b}}\Biggr) \frac{M_{\bullet}^2}{(M_{p} + M_{\bullet})^{1/3}} \chi \sin\lambda_{\bullet}\sin\nu_{\bullet} \,.
\ee 
Here $M_{\bullet}$ is the black hole mass, $M_{p}$ is the pulsar mass, $P_{b}$ is the orbital period of the binary system and $\chi$ is the dimensionless spin of the black hole. For the exact definition of the angles $\lambda_{\bullet} ,\nu_{\bullet}$, that describe the orientation of the spin of the black hole, we refer to figure 2 in the paper \cite{Wex1999}.
When $M_{p}<<M_{\bullet}$, the following numerical estimate applies for the constant factor
\be
T_{\odot}^{5/3} \Biggl(\frac{2\pi}{P_{b}}\Biggr) \frac{M_{\bullet}^2}{(M_{p} + M_{\bullet})^{1/3}} \approx (0.0001 \mu s) \Biggl(\frac{P_{b}}{1 \text{day}}\Biggr)^{-2/3} \Biggl(\frac{M_{\bullet}}{10}\Biggr)^{5/3}
\ee 

For the case of an edge-on circular equatorial orbit of the pulsar we have $\textit{i} = \nu_{\bullet}=\lambda_{\bullet}=\pi/2$, implying $\textit{A}_{FD} = 0$. We set $u = \varphi - \pi/2$ later in order to compare this formula with other approaches.

\subsubsection{Rafikov-Lai time delay}
In \cite{Rafikov2005}, Rafikov and Lai included the effect of the gravitational light bending into account, specially for a binary pulsar with highly-inclined orbit. The frame dragging delay is then given by 
\be\label{eq:R-L} 
\Delta t_{\text{FD}} = - (\Delta t_{\text{FD}}^{\text{max}}) a \sin\lambda_{\bullet} \Biggl(\frac{R_{E}}{b_{\pm}} \Biggr) \frac{\sin\nu_{\bullet}\cos u - \cos\nu_{\bullet}\sin u\cos\textit{i}}{(1-\sin^2 u\sin^2\textit{i})^{1/2}}
\ee 
where a fiducial unit of time is introduced 
\be 
(\Delta t)_{\text{FD}}^{\text{max}} \approx 1.44 \times 10^{-2} \mu s \Biggl( \frac{M_{c}}{M_{\odot}}\Biggr)^{3/2} \Biggl(\frac{R_{\odot}}{a_{\parallel}}\Biggr) 
\ee 
where $b_{\pm}$ is the impact parameters corresponding to the position of the two images,  $R_{E}$ is the Einstein radius and $a_{\parallel} = a\sin\textit{i} \sqrt{1-e^2}/(1+e\sin\omega)$ .
In the limit of $R\gg R_{E}$ equation (\ref{eq:R-L}) reduces to (\ref{eq:W-K}) where the lensing effect is neglected. 

We note that the Post-Newtonian delays reviewed here are in general coordinate dependent quantities what has to be considered for a comparison.

\section{Comparison to weak field approximations}
\begin{figure}
	\centering
	\includegraphics[width=0.35\textwidth]{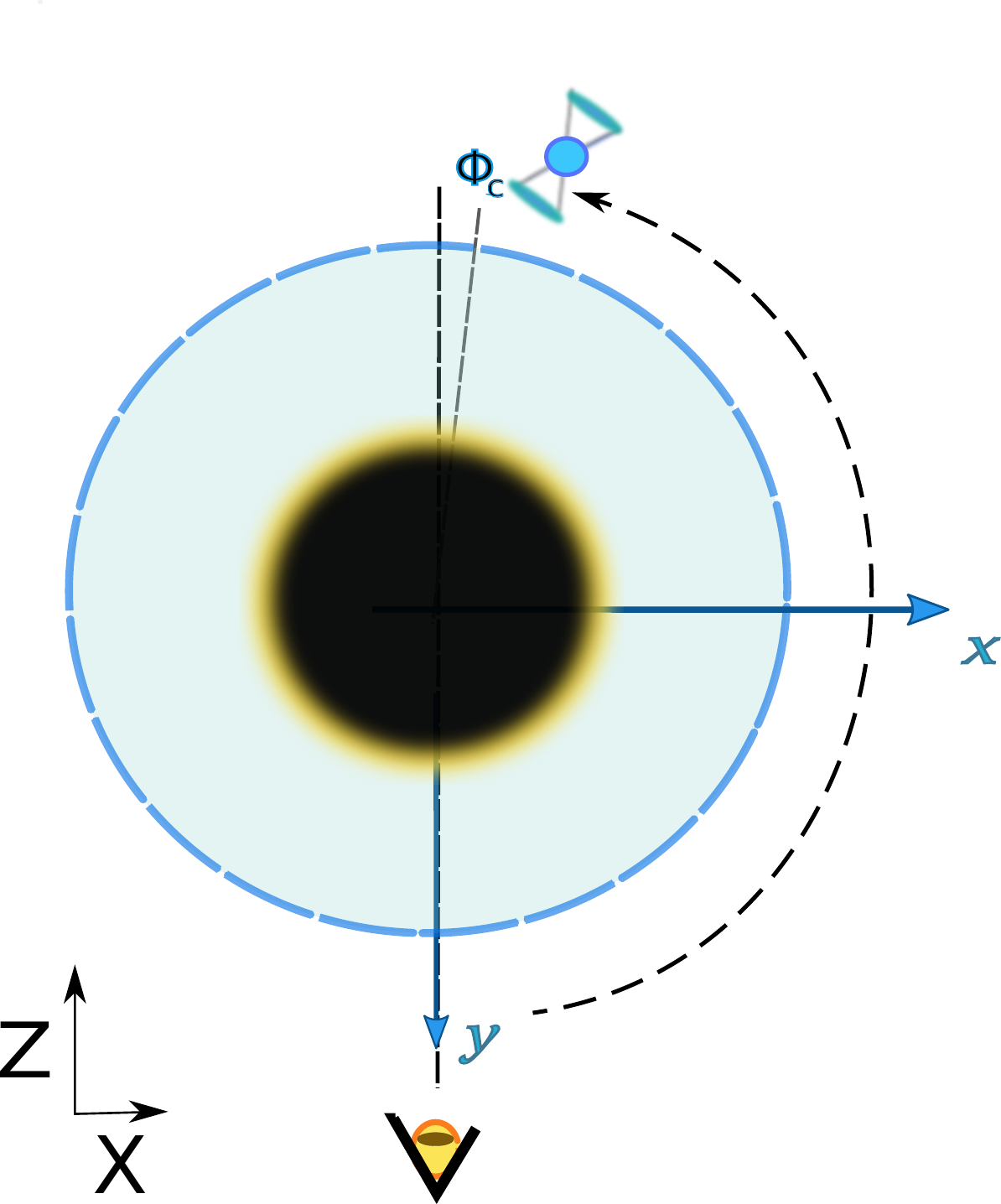}
	\caption{Illustration of the edge-on example below, where  the  pulsar  as well as  the observer are located in the equatorial plane identified with the galactic plane. $\phi_{c}< \pi$, denotes the critical angle at which primary lightlike geodesics make the switch from  counter to corotation with the respect to the spinning black hole in order to reach the observer at a shorter time.}
	\label{fig:diagram2}
\end{figure}
In this section, we compare our derived exact formula for the propagation time delay in a Kerr spacetime with both the formula derived for Schwarzschild spacetime in \cite{Hackmann2019} and the post-Newtonian approximations reviewed in section \ref{sec:weakfield}. 

We assume an extreme binary system of a pulsar, treated as a test particle, orbiting a supermassive black hole with a mass of $M = 4\times 10^6 M_{\odot}$ (solar masses), where $GM_{\odot}/c^2 =1476\,\rm M$. The propagation time delay will be expressed in seconds, therefore the dimensionless value can be recovered by dividing by $GM/c^3 \approx 19.7s$. The results can be rescaled for a different black hole mass $M_{2}$ by a factor $M_{2}/M$.

In the following we first analyse the complete formula given in equation \eqref{eq:prop_delay} before focusing specifically on the frame dragging effect defined in \eqref{eq:FD-Full} and its weak field counterparts reviewed in the foregoing section.

\subsection{Discussion of the complete propagation time delay}

\begin{figure}
	 \centering
	\includegraphics[width=0.48\textwidth]{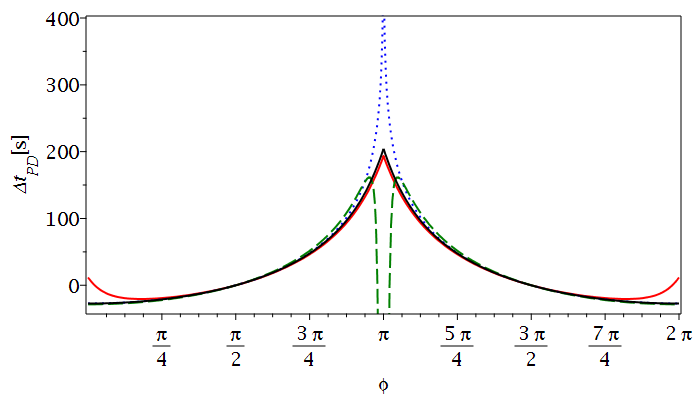}
	\includegraphics[width=0.48\textwidth]{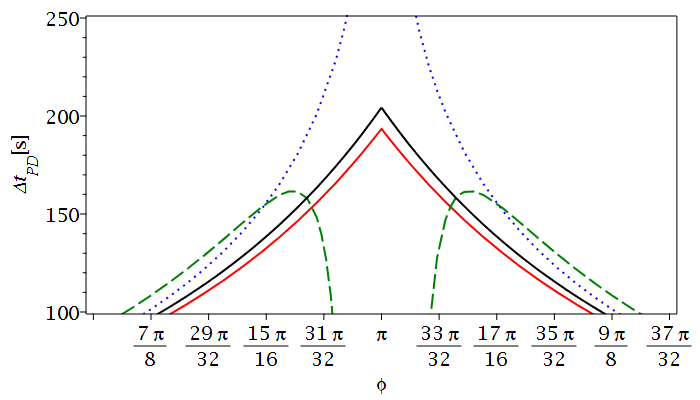}
	\caption{Schwarzschild Black hole with $a\approx0$:  the exact time delay subtracted by the Roemer delay  $\Delta t_{\text{ex}}- \Delta t_{\text{R}}$ (black line), compared with the first order Shapiro delay $\Delta t_{\text{S}}$ (dotted blue line), the lensed Shapiro delay including the geometric delay $\Delta t_{\text{S,l}} + \Delta t_{\text{geo}}$ (red line) and the second order delay $\Delta t_{\text{S}} + \Delta t_{2\text{PN}}$ (dash green line) for a circular edge-on orbit with a Schwarzschild radius $r_{\text{S}}=100\, M$. The bottom figure is a zoom of the top.}	 
	\label{fig:PN-delay-compare}
\end{figure}

Firstly, we test our main formula \eqref{eq:prop_delay} by choosing $a = 10^{-10}$ . For this value of the rotation $a$ the frame dragging effects are negligible, and we were able to reproduce the results from \cite{Hackmann2019}. For illustrative purposes, and to lay a ground for the discussions below, we show a single result in figure \ref{fig:PN-delay-compare}. For a detailed discussion of results for the non rotating case we refer to \cite{Hackmann2019}.

Note that the post-Newtonian expressions reviewed in section \ref{sec:weakfield} are given in harmonic coordinates. These are related to the coordinates of the Schwarzschild metric by $r_{\text{PN}} = r_{\text{Sch}} - M$, where $r_{\text{Sch}}$ is the usual Boyer-Lindquist type radial coordinate of the Schwarzschild metric.

In figure \ref{fig:PN-delay-compare} we show the time delay $\Delta t_{\text{ex}} - \Delta t_R$ (see equations \eqref{eq:prop_delay},\eqref{eq:Roemer}) together with the weak field approximations as discussed in section \ref{sec:weakfield} for a simple choice of the pulsar trajectory, an edge-on ($\textit{i}=\pi/2$) circular orbit. The ascending node with respect to the plane of sky is used as the reference point i.e. $\varphi_{\text{ref}} = \pi/2$, which with $\omega = - \pi/2$ simplifies to $\varphi_{\text{ref}} =\phi_{\text{ref}}= \pi/2$ (see \ref{eq:phi}). In this case, both the Roemer delay $\Delta t_{\text{R}}$ and the usual first order Shapiro delay $\Delta t_{\text{S}}$ vanish at the reference point, but the modified lensed delay  $\Delta t_{\text{S,l}} + \Delta t_{\text{geo}}$ as well as the second order delay $\Delta t_{2\text{PN}}$ show small offsets.  Also, the exact propagation time delay $\Delta t_{\text{ex}}$ includes a considerable offset. By adding global constants to the individual delays, we correct all these offsets such that they exactly vanish at $\phi_{\text{ref}} = \pi/2$. 

Let us point out some general features in figure \ref{fig:PN-delay-compare}. First, as expected for a circular edge-on orbit in spherical symmetry, the plot is symmetric with respect to the superior conjunction at $\phi=\pi$. If we switch on a rotation later on, this is expected to change, see below. Second, the usual Shapiro delays, both first and second order, diverge at $\phi=\pi$. This is related to the approximation of the light ray following a straight path, which then runs through the singularity. Finally, the lensed Shapiro delay, that avoids integrating through the singularity by using the lensed images as sources, provides a much better approximation around superior conjunction, but systematically underestimates the actual time delay. As the lensed delay is designed for superior conjunction it should not be used around inferior conjunction, where it indeed considerably deviates from both the exact delay and the usual Shapiro delay.



Now we turn to the case of a rotating black hole with spin parameter $a=0.9$. As already pointed out in section \ref{subsec:FD}, the Boyer-Lindquist coordinates used in our main formula \eqref{eq:prop_delay} do not have an intrinsic physical meaning. To compare results defined in different spcaetimes, we chose to identify the circumference of a circle. This then implied the relation \eqref{eq:rSK}. In the first order post-Newtonian metric, the circumference of a circular orbit of a radius $r_{\text{PN}}$, is $2\pi r_{\text{PN}}\sqrt{1 + 2 M/r_{\text{PN}}}$. Once the circumference is fixed, the relation $r_{\text{PN}} = \sqrt{M^2+r_{\text{Sch}}^2} - M$ holds. In the limit of large radii this reduces to $r_{\text{PN}} = r_{\text{Sch}} - M$.
Note that for simplicity, we denote in the following the Kerr radius $r_{\text{Kerr}}$ by $r$.


\begin{figure}
    \centering
    \includegraphics[width=0.48\textwidth]{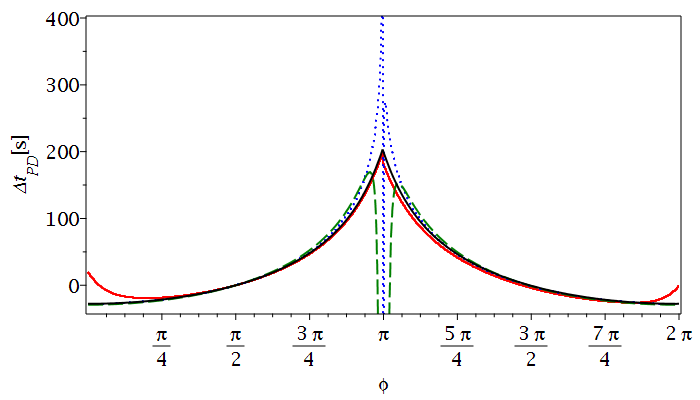}
    \includegraphics[width=0.48\textwidth]{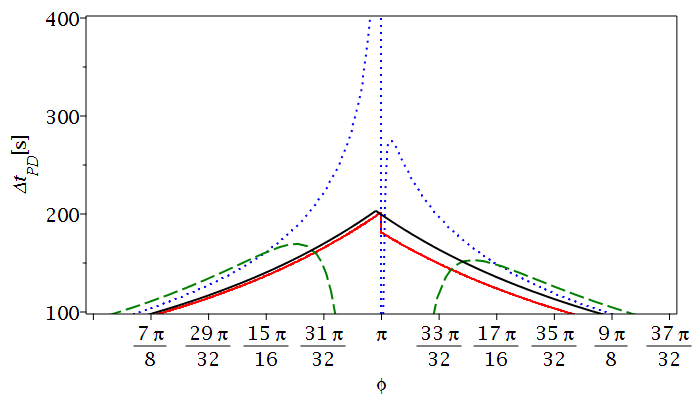}
    \caption{Rotating black hole $a=0.9$:  the exact Shapiro time delay subtracted by Roemer delay  $\Delta t_{\text{exact}}- \Delta t_{\text{R}}$(black line), compared with the first order Shapiro delay $\Delta t_{\text{S}}+\Delta t_{\text{WK}}$ (dotted blue line), the lensed Shapiro delay including the geometric delay $\Delta t_{\text{S,l}} + \Delta t_{\text{geo}}+\Delta t_{\text{RL}}$(red line) and the second order delay $\Delta t_{\text{S}} + \Delta t_{2\text{PN}}+\Delta t_{\text{WK}}$(dash green line) for an edge-on pulsar trajectory. Top, a circular orbit with a radius $r=100\,M$ (just to clearly display the effects we chose a unrealistically small radius) and bottom, a zoom of it}
    \label{fig:PD-FD-compare-1}
\end{figure}

In addition to the above discussed Post-Newtonian time delay expressions, we now consider \FD \ time delay terms in the weak field approximation, in particular the formula of Wex-Kopeikin and Rafikov-Lai (see \ref{sec:FD}).

As a representative example, we again chose a circular edge-on pulsar orbit with $r=100\,M$ (see figure \ref{fig:PD-FD-compare-1}). We immediately notice that, as expected and contrary to the spherically symmetric Schwarzschild case, the exact approach shows a slight shift in the angle of the maximum propagation time delay that correspond to the switch from contra-rotating to  co-rotating lightlike geodesics emitted from the pulsar to the observer. We denote the angle where this switch happens as $\phi_{c}$. In Kerr spacetime where the symmetry is only axial, we note that this switch happens early at $\phi_{c}<\pi$. This can be explained that until $\phi < \phi_{c}$, contra-rotating lightlike geodesics which are identified as primary light bundles, reach the observer earlier than the co-rotating ones that have longer trajectories. However, due to the spinning effect of the black hole on spacetime, some of these co-rotating emitted light signal from the pulsar at $\phi < \pi$ succeed to reach the observer quicker than the contra-rotating ones. This explains why we see $\phi_{c}<\pi$. As we expect, the latter phase shift $\phi_{c}$ converges toward $\pi$ for a larger pulsar orbit around the black hole.  This phenomena is one of the GR's predictions that is expected to be observable in such an extreme binary system.

\begin{table}
  \centering
\begin{tabular}{c c c c}
 & a = 0.1 & a = 0.5 & a = 0.9  \\ 
\hline
\hline
$\Delta t$ [$\mu s$] & 363.7  & 1065.8 & 2971.9 \\ 

\hline

$\Delta_{\phi}$ [$10^{-2} arcsec]$ & 0.3618 & 1.8086 & 3.2562  \\
\end{tabular}
\caption{For circular orbit with a radius $r = 10000\,M$ ($P\approx 3.92y$), the differences of the maximal exact propagation time delays between Kerr and Schwarzschild spacetime $\Delta t =|\text{max}(\Delta t _{PD,\text{Kerr}}) - \text{max}(\Delta t _{PD,\text{Schw}})|$ as well as  the corresponding  phases $\Delta_{\phi} = \pi - \phi_{c}$ .}
\label{tab:1}
\end{table}

The actual estimation for the spin of \SgrA is still speculative. Currently, three values are reported based on three different astrophysical techniques: $a = 0.1, 0.5, 0.9$ \cite{Fragione2020}, \cite{Genzel2003}, \cite{Witzel2018}. In table \ref{tab:1} we show for these three values of the spin,  the time difference between the maximal propagation time delays $\text{max}(\Delta t_{\text{PD,Kerr}})-\text{max}(\Delta t_{\text{PD,Schw}})$ of a Kerr and a Schwarzschild black hole. As just discussed, the maximal delays occur at different orbital phases in Kerr and Schwarzschild spacetime. The corresponding phase shift $\Delta_\phi := \pi - \phi_{c}$ for a fixed circular orbit with an orbital period $P\approx 3.92\,y$ is as well shown in table \ref{tab:1}. 

\begin{figure}
    \centering
    \includegraphics[width=0.48\textwidth]{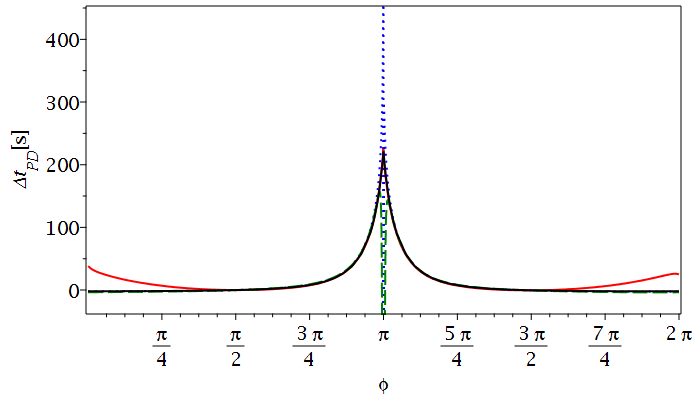}
    \includegraphics[width=0.48\textwidth]{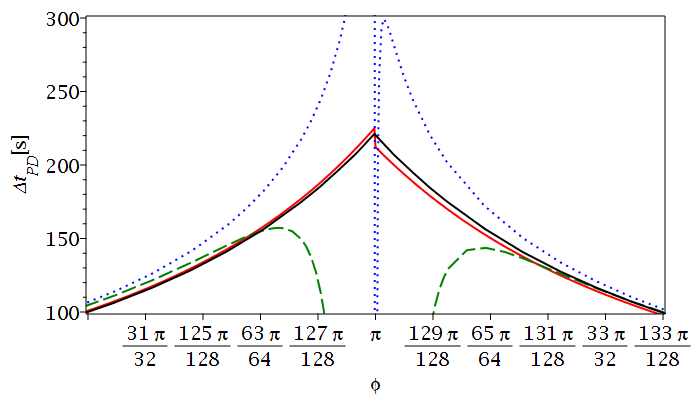}
    \caption{Rotating black hole $a=0.9$ with eccentricity $e=0.9$ and semi-major axis $A = 1000\,M$ ($P\approx 0.12 y$).The color of lines are identical to the previous figure. }
    \label{fig:PD-FD-compare-2}
\end{figure}

The shift of the maximal propagation delay is not reproduced by the post-Newtonian approximation considered here, even if we include the frame dragging terms due to Wex-Kopeikin or Rafikov-Lai, as can be seen in figures \ref{fig:PD-FD-compare-1} and \ref{fig:PD-FD-compare-2}. For the Rafikov-Lai term, we notice that it provides - apart from the missing shift of the maximum - a very good approximation of the effect, but still systematically underestimates it, in particular for $\phi>\pi$. In addition, it shows a rather unsatisfactory jump in the delay around superior conjunction $\phi=\pi$, which will be discussed in the next subsection. For the Wex-Kopeikin term, we still have the singularity at $\phi=\pi$, but now it diverges to $-\infty$ for $\phi > \pi$ in the first order. In that region the delay also has a local maximum to come back to the correct slope. Generally the time delay is overestimated by this approximation. Concerning the second order, away from superior conjunction it coincides well with our formula but fails to reproduce it around $\phi=\pi$. As expected, here the delay due to Wex-Kopeikin is not able to cancel the singularity.

Before we turn to discussing specifically the frame dragging delay terms in the next subsection, we drop the assumption of a circular orbit and consider as an example the case of a highly eccentric orbit with $e=0.9$. We assume a semi major axis $A = 1000\,M$ that is aligned with the observer-black hole line so that the pericentre is in front of the observer and the apocenter is exactly behind the black hole. The result is presented in figure \ref{fig:PD-FD-compare-2}. Note that the radial range of the pulsar orbit is between the apocenter at a distance  $d = 1900\,M$ and  pericenter at  $d = 100\,M$. We recognize basically the same features previously discussed for the circular orbit. Close to superior conjunction, we notice that the lensed delays, including the term by Rafikov-Lai (red line) now slightly overestimates the effect for $\phi<\pi$, before again it drops below the result of this paper (black line).
 
This reflects the change of the shape of the orbit due to extreme eccentricity. 
In passing we note that, comparing figure \ref{fig:PD-FD-compare-2} with figure \ref{fig:PD-FD-compare-1}, we notice also a slight deviation of the lensed delays (red line) with  our result (black line) close to the inferior conjunction due to the extreme eccentricity. However, as noted before, the lensed delays are designed to be used around superior conjunction and should anyway not be used around inferior conjunction.

\subsection{Discussion of the frame dragging time delay}
Now we will analyse the delays related to frame dragging. In the case of the exact delay, we have to stress that by comparing the expression in Kerr and Schwarzschild spacetime, in addition to the need to adjust for the different coordinates, we also automatically include effects that are caused by the different multipole structure of the Kerr black hole. In the post-Newtonian approach these effects are usually not explicitly included and have to be added by hand.


\begin{figure}
	 \centering
	\includegraphics[width=0.48\textwidth]{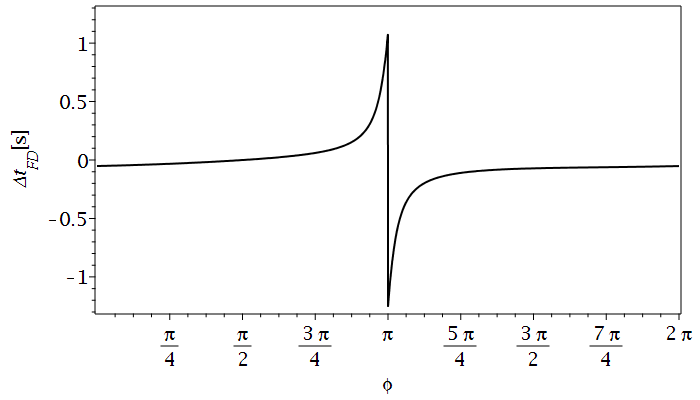}
    \includegraphics[width=0.48\textwidth]{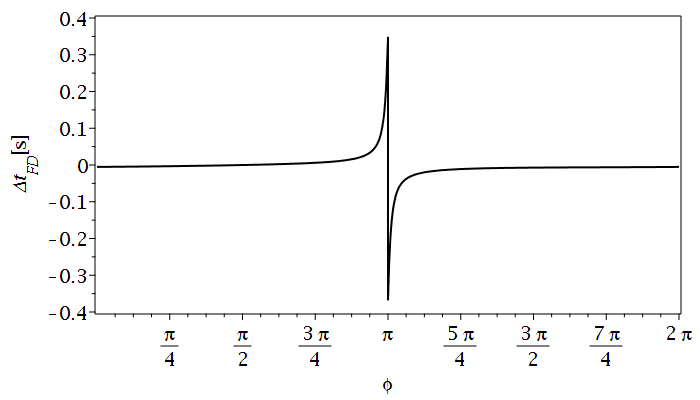}
	\caption{Relativistic Frame Dragging time delay for a rotating black hole with $a = 0.9$ and circular orbit. Top,  $r=1000\,M $, bottom $r = 10000\,M$.}	 
	\label{fig:FD-alone1}
\end{figure}

In figure \ref{fig:FD-alone1}, we plot for the case of a rotating black hole with $a = 0.9$, the relativistic exact \FD \ delay $\Delta t_{\text{FD}} = \Delta t_{\text{ex}}(a=0.9) - \Delta t_{\text{ex}}(a=0)$ for two choices of circular orbit $r=1000\,M, 10000\,M$.
We notice that in general there appears a maximum located shortly before $\phi=\pi$, exactly at the critical position $\phi_c$ where the signals switch from contra- to co-rotating trajectories. Then the curve sharply drops to a minimum, exactly at $\phi=\pi$, where again both for Schwarzschild and Kerr spacetime the primary light rays are co-rotating. With a larger radius, the shape of the relativistic \FD\ delay becomes sharper. This can be explained by the critical orbital position $\phi_{c}$ converging towards $\phi = \pi$ with increasing orbital radius of the pulsar.

\begin{figure}
	 \centering
	\includegraphics[width=0.48\textwidth]{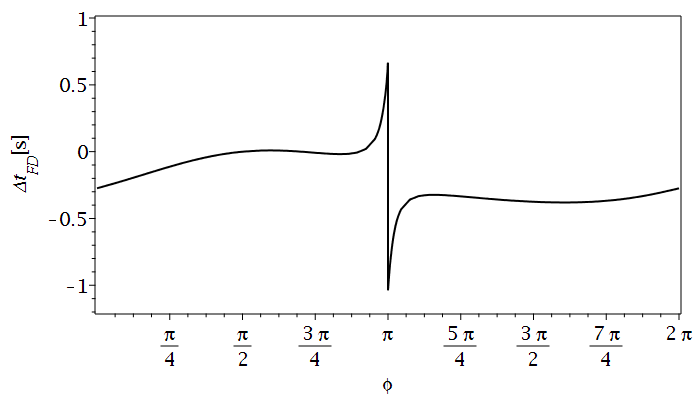}
    \includegraphics[width=0.48\textwidth]{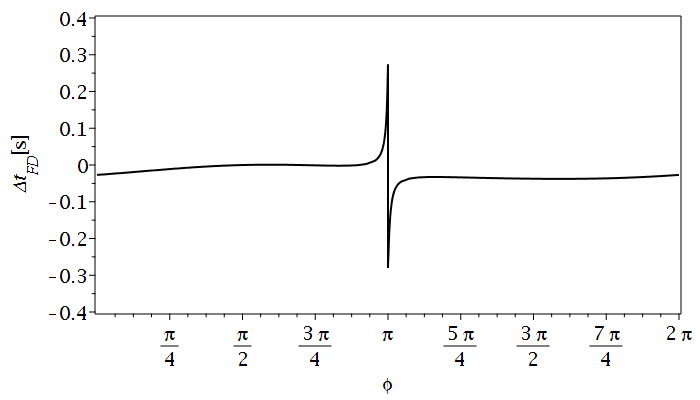}
	\caption{Relativistic Frame Dragging time delay for a rotating black hole with $a = 0.9$ and extreme eccentric orbit $e=0.9$. Top, semi-major axis $A=1000\,M$, bottom $A=10000\,M.$}	 
	\label{fig:FD-alone2}
\end{figure}
In figure \ref{fig:FD-alone2}, we plot the relativistic exact \FD \ delays  for the same choices of the semi major axis $A=1000\,M, 10000\,M$ with eccentricity $e = 0.9$. We notice that like in the circular orbit, the \FD effect reduces with a larger orbit around the black hole. However  the curve of the relativistic \FD delay changes its shape reflecting the eccentricity of the pulsar trajectory with its semi major axis oriented along the black hole-observer line.


\begin{figure}
	 \centering
	\includegraphics[width=0.48\textwidth]{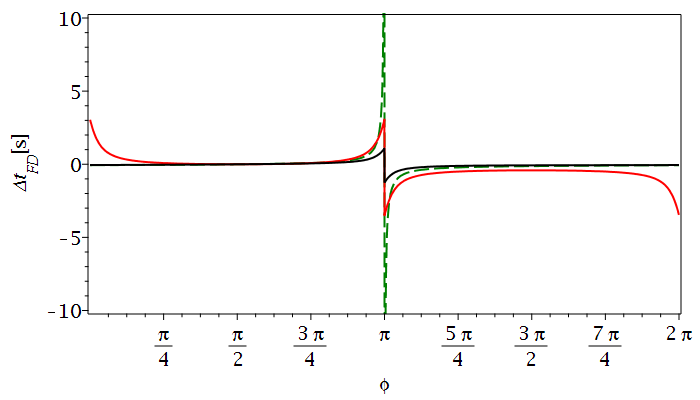}
    \includegraphics[width=0.48\textwidth]{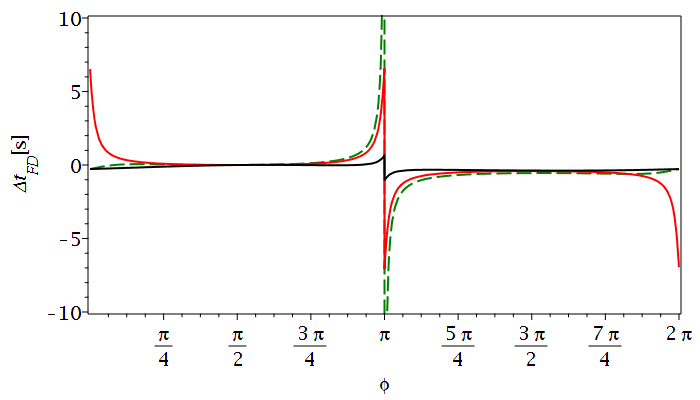}
\caption{Comparison of Frame Dragging delay between the relativistic (solid black line) and the Post-Newtonian approach of Wex-Kopeikin(green line) and Rafikov-Lai (red dashed line) for the elliptic edge-on orbits with eccentricity $e=0.9$ and with different semi major axis lengths. Top, a circular orbit with $r=1000\, M$, bottom an extreme elliptic orbit with $e =0.9$ and semi major-axis $A = 1000\,M$}	 
	\label{fig:FD-compare}
\end{figure}

Now we will compare the two post-Newtonian based approaches mentioned before for the \FD \ delays  with the relativistic exact formula derived in this paper. Figure \ref{fig:FD-compare} shows the \FD~ time delay as a function of the orbital phase of the pulsar around SgrA* for a circular as well as an extreme elliptic orbit with $r=A=1000\,M$. First, we notice that, while the Wex-Kopeikin approach shows a singularity at $\phi=\pi$, the formula by Rafikov-Lai keeps finite values along the orbit. This is again related to the chosen approximation of the lightlike trajectory, which for the Wex-Kopeikin approach passes through the singularity at superior conjunction, but not for the Rafikov-Lai formula. Moreover, from the equations for the Wex-Kopeikin and the Rafikov-Lai formula, it is clear that the delays behaves symmetric with respect to $\phi=\pi$ and do not reflect the phase shift of the switch from contra- to co-rotating trajectories discussed before, that is to be expected around a spinning black hole and is present in the expression derived in this paper.

At $\phi=0,\ 2\pi$ where the pulsar is in the front of the supermassive black hole, at pericenter, the Wex-Kopeikin formula coincides with our solution, whereas the \FD \ effect in Rafikov-Lai expression shows a maximal and a minimal value far away from the delay derived in this paper. This again emphasizes that the approach using lensed positions is designed for superior conjunction and should not be used around inferior conjunction.

\begin{figure}
	 \centering
	\includegraphics[width=0.48\textwidth]{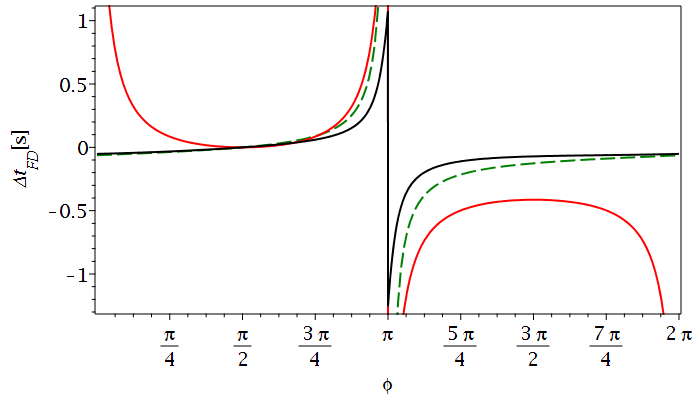}
    \includegraphics[width=0.48\textwidth]{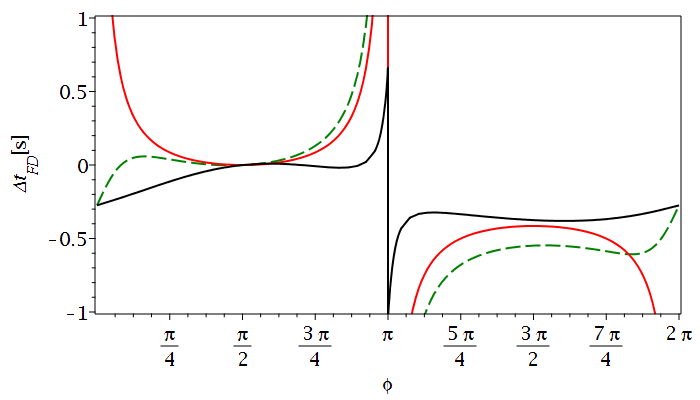}
\caption{Zoom in \ref{fig:FD-compare}}	 
	\label{fig:FD-compare-zoom}
\end{figure}

Generally, it can be seen that (in absolute values) the frame dragging effects are overestimated by the post-Newtonian formulas. For larger values of the orbital radius, the Wex-Kopeikin and Rafikov-Lai formula, however, converge toward our result,  as expected. We should mention here again that our formula does not only model the pure classical \FD effect but also the quadrupole moment  and all higher multipole moment effects that are induced by the spin of \SgrA \ as compared to spherical symmetry.

\section{Disturbing effects}
We expect that it would be hard to measure directly the \FD ~ time delay since other effects could screen it partially or totally. Therefore it is worth to mention these effects bellow.  

\subsection{Relativistic Doppler Shift of the pulsar}
One of these effects is the Doppler shift of the pulsar along its trajectory. Along the trajectory $0 \le \varphi \le \pi$ the pulsar is moving  away from the observer and therefore the emitted light would be red-shifted, whereas for the trajectory $\pi\le \varphi \le 2\pi$ the pulsar is moving toward the observer and therefore the light ray becomes  blue-shifted.
It is expected that a Pulsar located at the vicinity of SgrA$^{*}$ would have a considerable high velocity. The S2, also known as S0–2, is a star that is located in the star cluster close to SgrA$^{*}$, orbiting it with a period of $16.0518$ years. Approaching the epicenter the speed of S2  is expected to exceed 5,000 km/s which represents a $1/60$ the speed of light. A pulsar in the near location would have the same order of magnitude therefore we expect that the relativistic Doppler Shift effect of the pulsar would be significant. Based on special relativity theory the Doppler Shift is given by 
\be\label{eq:doppler-shift} 
z = 1 \pm \frac{v}{c}
\ee 
where $v$ is the velocity of the pulsar, the plus and minus sign correspond consequently to a pulsar moving away and toward the observer. 
For a fixed observing frequency, the blue shifted photon with a frequency $f_{1}$ and the other red shifted one with a frequency $f_{2}$ will induce a time delay based on the dispersion measure according to the formula (4.7) in \cite{Lorimer2005}:
\be\label{eq:DM-t} 
\Delta t_{\text{Doppler}} \approx 4.15 \times 10^6 \text{ms}  \times (f_{1}^{-2} - f_{2}^{-2}) \times \text{DM}
\ee 
For the DM value we choose $\text{DM} = 1778 \text{cm}^{-3} \text{pc}$ of the nearest found magnetar J1745-2900 to SgrA$^{*}$

In table \ref{tab:2} we listed for different radii the estimated Doppler-Shift delay for a fixed light frequency of $2500\,\text{MHZ}$.  

\subsection{Gravitational red shift}
Due to the high mass of the super-massive black hole SgrA$^{*}$, the gravitational red-shift must  be taken into account which could modify the measured frame dragging delay from the predicted theoretical one. The red shift factor could be estimated based on  \cite{Gates2021}
\be\label{Gates-Hadar-Lupsasca}
g_{\pm} = \frac{\sqrt{r^3-3 M r^2\pm 2a\sqrt{M}r^{3/2}}}{r^{3/2}\pm \sqrt{M}(a-\lambda)}
\ee 
where the plus and minus sign corresponds to co and contra-rotating light-like geodesics respectively. With the help of (\ref{eq:DM-t}) the corresponding gravitational red shift delay $\Delta t_{\text{grav-shift}}$ is given in table \ref{tab:2} with the same choice of radii for a circular pulsar orbit around \SgrA.  

\subsection{Frame dragging effect on the pulsar orbit}
The frame dragging will not only effect the photon trajectory from the pulsar to the observer but also the pulsar as an emission point. Therefore it is worth to consider this effect on the pulsar orbit. The additional velocity due to frame dragging can be approximated by the ratio of the excess in the circumference of the circular orbit in Kerr spacetime divided by its orbital period 
\be 
\Delta_{v} = \frac{2\pi r (\sqrt{1+\frac{a^2}{r^2}+\frac{2Ma^2}{r^3}}-1)}{P_{b}}
\ee 
Based on (\ref{eq:doppler-shift}) and (\ref{eq:DM-t}) the corresponding time delay $\Delta t_{\text{FD,pulsar}}$ is calculated in table \ref{tab:2} for the same choice of orbital radii.  
 
\begin{table}
  \centering
\begin{tabular}{c c c c c}
          & $r=10^2\,M$  &$ r=10^3\,M$ & $r=10^4\,M$ \\
 &$P\approx 1.41d$  & $P\approx 0.12y$ &$ P\approx 3.92y$  \\ [0.2ex] 
\hline
\hline
$\Delta t_{\text{rel. FD}} [s]$ & 8.025  & 2.974  & 0.717  \\
\hline 
$\Delta t_{\text{rel. Doppler}}[s]$ & 0.4897 & 0.1547 & 0.04728 \\ 
\hline
$\Delta t_{\text{grav-shift}}[s]$ & 7.17027$\times10^{-3}$ & 2.11557$\times10^{-4}$ & 6.687$\times10^{-6}$  \\
\hline
$\Delta t_{\text{FD,orbit}}[s]$ & 0.952$\times10^{-3}$  & 6.266$\times10^{-8}$ & 1.915$\times10^{-10}$  \\
\end{tabular}
\caption{relativisitc \FD time delays compared with relativistic Doppler effect $\Delta t_{\text{Doppler}}$, gravitational red shift $\Delta t_{\text{Grav-shift}}$ and Frame Dragging effect on the pulsar's motion $\Delta t_{\text{FD,pulsar}}$ induced time delays for circular edge on orbits with different radii $r$ and a choice for the black hole spin $a=0.9$}
\label{tab:2}
\end{table}

In order to estimate their presence in the time of arrival of the pulsar's radio waves (ToAs), we calculate the previously discussed disturbing effects, in addition to the relativistic \FD delay, in table \ref{tab:2} for 3 choices of circular orbits. We notice that while the gravitational red shift as well as the frame dragging effect on the pulsar orbit may be detectable by the timing of a milli second pulsar, but they are still dominated by the the relativistic Doppler shift by a  minimum factor of 100 for a very close pulsar orbit ($r = 100\, M$). In general case, these latter effects, even  combined together,  can screen only partially the amplitude of \FD time delay by less than $10\%$ in the worst case, which makes it  measurable.

This is only right for the edge on case where \FD is maximal. But in a general case the conclusion above can be still valid: for an inclined pulsar orbit we can assume that the \FD delay scales by a factor of $\sin\textit{i}$, therefore  the amplitude of the \FD would be comparable with the rel. Doppler if $\textit{i} = \pi/50$  for $r = 100\,M$ which is possible but still far from what we expect.

We should mention another screening effect, the bending delay derived by \cite{Doroshenko1995}. It has the same functional dependence as the post-Newtonian frame dragging delay, and can hardly be disentangled from it. However, for the case considered here, namely a pulsar orbiting a supermassive black, the frame dragging effect should be much stronger and dominate the bending delay completely.

%
\section{Summary and outlook}
In this paper we derived an exact analytical solution for the time delay of lightlike geodesics in Kerr spacetime in terms of Jacobian elliptic integrals. By isolating the diverging parts we were able to find an explicit analytical formula for the finite propagation delay with respect to a reference point. This result can be interpreted as the relativistic propagation delay of the signals of pulsars orbiting a rotating supermassive black hole, where the extreme mass ratio justifies to consider the pulsar as a test particle. We then compared our result for the propagation delay in pulsar timing  to known post-Newtonian expressions. Moreover, we derived  from the difference between the propagation time delays in Kerr and Schwarzschild spacetimes  the exact relativistic \FD\, time delay. We adopted the harmonic coordinates in order to compare the time delays derived here with the ones in post-Newtonian spacetimes. As an illustration of our general results, we explicitly treated the case of equatorial motion, and edge-on pulsar orbits, where the propagation delay as well as the \FD delay are expected to be maximal. For this case our results showed that the post-Newtonian based treatments of \WK and \RL generally overestimate the frame dragging effects, in particular around and after superior conjunction. Based on this, we expect that our approach should be more reliable and accurate in extreme mass ratio binary configurations, where at some point the pulsar orbit will be at or close to superior conjunction. Furthermore we presented some disturbing effects that could screen partially or fully  the pure \FD\ delay from timing the pulsar. \\
The advantage of this work, as compared to numerical ray tracing methods, is that the derived analytical expression could be integrated into a new relativistic timing model  that is suitable for such an extreme mass ratio binary system  and later processed as a plugin within the TEMPO2 pulsar timing package. To this end, a complementary continuation of this work is to identify the constants of motion $\lambda$ and $q$ of the lightlike geodesics with orbital parameters of the pulsar, possibly using analytical approximations in the Kerr field in analogy to similar works in Schwarzschild spacetime, see the review \cite{Semerak2015} and references therein. This would permit to use the derived  formula for propagation as well as frame dragging delay to be integrated within a post-Keplerian timing model similar to "ELL1" or "T2" orbital model used for binary pulsar system.\\
A natural continuation of our work would be to include other general relativity's spacetimes, in particular the most general type-D black hole solutions, the Pleba\'nski-Demia\'nski metric  with additional parameters like electric, magnetic and NUT charge and to investigate their corresponding effects on the propagation time delay. In principle, the equations of motion for the lightlike geodesics are solvable in terms of elliptic or hyperelliptic integrals. Another possibility would be to include parameterised spacetimes, as long as they allow for separable Hamilton-Jacobi equations. Such investigations would probe the possibility to test the No-Hair theorem by predicting the time delays induced by these additional parameters if the black hole has more hair than the mass and the spin.  
Another possible extension of the work is to treat the pulsar as a point particle with a spin and to study the possible coupling effects between the spin and the pulsar orbit as well as the spin and the curved spacetime around the supermassive black hole within the accuracy of timing a milli second pulsar. In the work of \cite{Kimpson2020}, the spin effects were treated numerically through light ray tracing method but without giving any explicit analytical expression for the corresponding time delay as a function of the orbital Kepler parameters or the emitted lightlike geodesics constants.\\
$\mathbf{Acknowledgements}$ The authors are thankful for Volker Perlick for his delightful comments on this paper. This project is supported by the research training group GRK 1620 ”Models of Gravity”, funded by the Deutsche Forschungsgemeinschaft (DFG, German Research Foundation). E.H. acknowledges the excellence cluster EXC-2123 QuantumFrontiers – 390837967 funded by the Deutsche Forschungsgemeinschaft (DFG, German Research Foundation) under Germany’s Excellence Strategy.\\

\newpage
~
\newpage


\setcounter{secnumdepth}{0}
\section{Appendix}\label{Appendix}
Here we  derive  the analytical solution of the (\ref{eq:Int-r-2})in terms of elliptic integrals. 
With $r_{\pm} = M \pm \sqrt{M^2-a^2}$ we the integral splits in a sum of simple integrals \\
\begin{align}
&\int \frac{r^2(r^2+a^2)+2Mar(a-\lambda)}{\Delta \sqrt{\R(r)}}dr=\int \frac{r^2}{\sqrt{\R(r)}} \nonumber \\
&+ 2M\int \frac{r}{\sqrt{\R(r)}}dr + 4M^2\int \frac{dr}{\sqrt{\R(r)}}\nonumber \\
&-\frac{M}{\sqrt{M^2-a2}}\Biggl[ ((a\lambda - 4M^2)r_{+} + 2Ma^2) \int \frac{dr}{(r-r_{+})\sqrt{\R(r)}} \\
&+(-(a\lambda - 4M^2)r_{-} + 2Ma^2) \int \frac{dr}{(r-r_{-})\sqrt{\R(r)}} \Biggr] \nonumber
\end{align}
 The substitution 
\be\label{eq:x} 
x^2 = \frac{(r-r_{4})(r_{3} - r_{1})}{(r-r_{3})(r_{4}-r_{1})} \label{eq:defx}
\ee
with the roots $r_{i}$ of $\R$ chosen as in (\ref{eq:R-roots}), casts the integrals in general  in the Legendre form
\begin{align}
T(r,\lambda)=\frac{2M}{(r_{3}-r_{1})(r_{4}-r_{2})}\biggl[\sum_{i} \int_{0} ^{x(r)} \frac{f_{i}(x)}{\sqrt{(1-x^2)(1-k^2x^2)}}\Biggr] \label{eq:k} 	 
\end{align}
where
\be
k^2=\frac{(r_{3}-r_{2})(r_{4} - r_{1})}{(r_{3}-r_{1})(r_{4}-r_{2})} 
\label{eq:defk}
\ee 
and $f_{i}(x)$ are rational functions. We define the following constants:
\be
\begin{aligned} 
c_{1}=\frac{r_{4} - r_{1}}{r_{3} - r_{1}}, ~ c_{2} = \frac{r_{3} - r_{2}}{r_{4}-r_{2}}, ~ c_{\pm}=\frac{(r_{3}- r_{\pm})(r_{1}-r_{4})}{(r_{4}-r_{\pm})(r_{1}-r_{3})} \label{eq:defc}
\end{aligned}
\ee  
We note that for $r = \infty$ (\ref{eq:x}) reduces to 
\be 
x^{2}_{\infty} = x(r=\infty)^2 = \frac{r_{4}-r_{2}}{r_{4}-r_{1}}
\ee 
With the Jacobi elliptic integrals 
\be 
F(x,k) &=& \int_{0} ^x \frac{dx}{\sqrt{(1-x^2)(1-k^2x^2)}} \\
E(x,k) &=& \int_{0}^{x} \sqrt{1-k^2x^2} dx \\
\Pi(x,c,k) &=& \int_{0}^{x} \frac{dx}{(1-cx^2)\sqrt{(1-x^2)(1-k^2x^2)}}
\ee 
we find 

\begin{align}\label{eq:Tr-1} 
&T(r,\lambda)=\frac{2M}{\sqrt{(r_{3}-r_{1})(r_{4}-r_{2})}} \Biggl[\Biggl(2(r_{3} +2M) + \frac{r_{3}r_{4}-r_{1}r_{4}+r_{1}r_{3}+r_{3}^2}{2M}\nonumber\\
&+\frac{(a\lambda - 4M^2)r_{-}-2Ma^2}{(r_{3}-r_{-})\sqrt{M^2-a^2}}-\frac{(a\lambda - 4M^2)r_{+}+2Ma^2}{(r_{3}-r_{+})\sqrt{M^2-a^2}}\Biggr) F(x,k)\nonumber \\
&-\frac{(r_{3}-r_{1})(r_{4}-r_{2})}{2M}E(x,k) + 2(r_{4}-r_{3})\Pi(x,c_{1},k)\\
&+ \frac{(r_{+}a\lambda - 4r_{+}M^2 + 2Ma^2)(r_{4}-r_{3})}{\sqrt{M^2-a^2}(r_{3}-r_{+})(r_{4}-r_{+}))}\Pi(x,c_{+},k)\nonumber\\
&-\frac{(r_{-}a\lambda - 4r_{-}M^2 - 2Ma^2)(r_{4}-r_{3})}{\sqrt{M^2-a^2}(r_{3}-r_{-})(r_{4}-r_{-}))}\Pi(x,c_{-},k)\Biggr] +\frac{\sqrt{\R}}{r-r_{3}} \nonumber
\end{align}

where $x$ is related  to $r$ via (\ref{eq:x}). Note that the Jacobi elliptic integrals can be evaluated without using a numeric integration, and can therefore be considered as an exact analytical solution to the integral $T$.
Note that the last term in (\ref{eq:Tr-1}) diverges linearly for $r \rightarrow \infty$. As well, $\Pi(x,c_{1},k)$ diverges logarithmic  for $x^2=1/c_{1}$, which happens in our case for $x = x_{\infty}$ and $c = c_{2}$. Therefore, the time for reaching $r = \infty$ diverges as expected. To isolate the diverging term we apply the following identity for $c=c_{1}$:  
\be 
\Pi(x,c,k) = F(x,k) - \Pi(x,\frac{k^2}{c},k) + \frac{\ln(Z)}{2P} 
\ee      
where 
\be 
Z &=& \frac{\sqrt{(1-x^2)(1-k^2x^2)} + Px}{\sqrt{(1-x^2)(1-k^2x^2)} - Px} \\
P^2 &=& \frac{(c-1)(c-k^2)}{c} = \frac{(r_{3}-r_{4})^2}{(r_{4}-r_{2})(r_{3}-r_{1})}
\ee 
With this equation (\ref{eq:Tr-1}) becomes 
\begin{align}\label{eq:Tr-2} 
 &T_{r}(\lambda,q) = \frac{2M}{\sqrt{(r_{3}-r_{1})(r_{4}-r_{2})}} \Biggl[\Biggl(4M + \frac{r_{3}r_{4}-r_{1}r_{4}+r_{1}r_{3}+r_{3}^{2}}{2M} \nonumber\\
&+ 2r_{4} + \frac{(a\lambda - 4M^{2})r_{-} - 2M a^{2}}{(r_{3}-r_{-})\sqrt{M^{2}-a^{2}}} 
- \frac{(a\lambda - 4M^2)r_{+} + 2Ma^2}{(r_{3}-r_{+})\sqrt{M^2-a^2}}\Biggr)F(x,k)\nonumber\\
&-\frac{(r_{3}-r_{1})(r_{4}-r_{2})}{2M}E(x,k)-2(r_{4}-r_{3})\Pi(x,c_{2},k) \nonumber\\
&+ \frac{(r_{4}-r_{3})}{\sqrt{M^2-a^2}} \Biggl(\frac{(r_{+}a\lambda-4r_{+}M^2+2Ma^2)}{(r_{3}-r_{+})(r_{4}-r_{+})} \Pi(x,c_{+},k)\\ 
&- \frac{(r_{-}a\lambda-4r_{-}M^2-2Ma^2)}{(r_{3}-r_{-})(r_{4}-r_{-})} \Pi(x,c_{-},k) \Biggr)\Biggr] \nonumber\\ 
&+ \frac{\sqrt{\R(r)}}{r-r_{3}} + 2M \ln(\frac{\sqrt{(r-r_{2})(r-r_{1})}+\sqrt{(r-r_{4})(r-r_{3})}}{\sqrt{(r-r_{2})(r-r_{2})}-\sqrt{(r-r_{4})(r-r_{3})}}) 
\end{align}
where the last two terms diverge for $x=x_{\infty}$. We find the Taylor expansion of these terms as 
\be 
\frac{\sqrt{\R(r)}}{r-r_{3}} &=& r + r_{3} + \mathcal{O}(\frac{1}{r}) \\
\ln(Z) &=&  \ln(\frac{2}{r_{3}+r_{4}}) + \ln(r) + \mathcal{O}(\frac{1}{r})   
\ee
\\
The analytical solution of the eq.(\ref{eq:int-u}) in terms of elliptic integrals is given by : 
\be
T(u,\lambda) &=& \frac{I~\text{sign}(a)}{\sqrt{u_{+} - u_{-}}} \Biggl( (u_{-} - u_{+}) E(v,w) + u_{+} F(v,w)\Biggr)\\
\text{where}\nonumber\\
v &=& \sqrt{\frac{u_{-}-u}{u_{-}}} \label{eq:defv}\\
w &=& \sqrt{\frac{u_{-}}{u_{-}-u_{+}}} \label{eq:defw}
\ee
and
\be 
u_{\pm} &=& \frac{a^{2}- \lambda^{2} - q \pm \sqrt{(a^{2}-\lambda^{2}-q)^{2} + 4 a^{2} q}}{2 a^{2}}
\ee 
are the solutions of the second order polynomial,
\be 
0 = q + u (a^2 - \lambda^2 - q) - a^2 u^2 = \frac{U_{\lambda,q}(u)}{u}
\ee 



\bibliographystyle{mnras}
\bibliography{bibliography}

\begin{thebibliography}{}
\makeatletter
\relax
\def\mn@urlcharsother{\let\do\@makeother \do\$\do\&\do\#\do\^\do\_\do\%\do\~}
\def\mn@doi{\begingroup\mn@urlcharsother \@ifnextchar [ {\mn@doi@}
  {\mn@doi@[]}}
\def\mn@doi@[#1]#2{\def\@tempa{#1}\ifx\@tempa\@empty \href
  {http://dx.doi.org/#2} {doi:#2}\else \href {http://dx.doi.org/#2} {#1}\fi
  \endgroup}
\def\mn@eprint#1#2{\mn@eprint@#1:#2::\@nil}
\def\mn@eprint@arXiv#1{\href {http://arxiv.org/abs/#1} {{\tt arXiv:#1}}}
\def\mn@eprint@dblp#1{\href {http://dblp.uni-trier.de/rec/bibtex/#1.xml}
  {dblp:#1}}
\def\mn@eprint@#1:#2:#3:#4\@nil{\def\@tempa {#1}\def\@tempb {#2}\def\@tempc
  {#3}\ifx \@tempc \@empty \let \@tempc \@tempb \let \@tempb \@tempa \fi \ifx
  \@tempb \@empty \def\@tempb {arXiv}\fi \@ifundefined
  {mn@eprint@\@tempb}{\@tempb:\@tempc}{\expandafter \expandafter \csname
  mn@eprint@\@tempb\endcsname \expandafter{\@tempc}}}

\bibitem[\protect\citeauthoryear{{Blandford} \& {Teukolsky}}{{Blandford} \&
  {Teukolsky}}{1976}]{Blandford1976}
{Blandford} R.,  {Teukolsky} S.~A.,  1976, \mn@doi [Astrophys. J.]
  {10.1086/154315}, \href {http://adsabs.harvard.edu/abs/1976ApJ...205..580B}
  {205, 580}

\bibitem[\protect\citeauthoryear{{Broderick}, {Johannsen}, {Loeb}  \&
  {Psaltis}}{{Broderick} et~al.}{2014}]{Broderick2014}
{Broderick} A.~E.,  {Johannsen} T.,  {Loeb} A.,   {Psaltis} D.,  2014, \mn@doi
  [\apj] {10.1088/0004-637X/784/1/7}, \href
  {http://adsabs.harvard.edu/abs/2014ApJ...784....7B} {784, 7}

\bibitem[\protect\citeauthoryear{Carter}{Carter}{1968}]{Carter1974}
Carter B.,  1968, \mn@doi [Phys. Rev.] {10.1103/PhysRev.174.1559}, 174, 1559

\bibitem[\protect\citeauthoryear{Chandrasekhar}{Chandrasekhar}{1998}]{Chandrasekhar1998}
Chandrasekhar S.,  1998, The Mathematical Theory of Black Holes.
International series of monographs on physics, Clarendon Press, \url
  {https://books.google.de/books?id=LBOVcrzFfhsC}

\bibitem[\protect\citeauthoryear{Christian, Psaltis  \& Loeb}{Christian
  et~al.}{2015}]{Christian2015}
Christian P.,  Psaltis D.,   Loeb A.,  2015, arXiv: General Relativity and
  Quantum Cosmology

\bibitem[\protect\citeauthoryear{Collaboration}{Collaboration}{2019a}]{EHT2019i}
Collaboration T. E. H.~T.,  2019a, The Astrophysical Journal Letters, 875

\bibitem[\protect\citeauthoryear{Collaboration}{Collaboration}{2019b}]{EHT2019ii}
Collaboration T. E. H.~T.,  2019b, The Astrophysical Journal Letters, 875

\bibitem[\protect\citeauthoryear{Collaboration}{Collaboration}{2019c}]{EHT2019iii}
Collaboration T. E. H.~T.,  2019c, The Astrophysical Journal Letters, 875

\bibitem[\protect\citeauthoryear{Collaboration}{Collaboration}{2019d}]{EHT2019vi}
Collaboration T. E. H.~T.,  2019d, The Astrophysical Journal Letters, 875

\bibitem[\protect\citeauthoryear{{Cunha}, {Font}, {Herdeiro}, {Radu},
  {Sanchis-Gual}  \& {Zilh{\~a}o}}{{Cunha} et~al.}{2017}]{Cunha2017}
{Cunha} P.~V.~P.,  {Font} J.~A.,  {Herdeiro} C.,  {Radu} E.,  {Sanchis-Gual}
  N.,   {Zilh{\~a}o} M.,  2017, \mn@doi [\prd] {10.1103/PhysRevD.96.104040},
  \href {http://adsabs.harvard.edu/abs/2017PhRvD..96j4040C} {96, 104040}

\bibitem[\protect\citeauthoryear{{Damour} \& {Deruelle}}{{Damour} \&
  {Deruelle}}{1986}]{Damour1986}
{Damour} T.,  {Deruelle} N.,  1986, Ann.~Inst.~Henri Poincar{\'e}
  Phys.~Th{\'e}or., Vol.~44, No.~3, p.~263 - 292, \href
  {http://adsabs.harvard.edu/abs/1986AIHS...44..263D} {44, 263}

\bibitem[\protect\citeauthoryear{{Damour} \& {Taylor}}{{Damour} \&
  {Taylor}}{1992}]{Damour1992}
{Damour} T.,  {Taylor} J.~H.,  1992, \mn@doi [\prd] {10.1103/PhysRevD.45.1840},
  \href {http://adsabs.harvard.edu/abs/1992PhRvD..45.1840D} {45, 1840}

\bibitem[\protect\citeauthoryear{{Doroshenko} \& {Kopeikin}}{{Doroshenko} \&
  {Kopeikin}}{1995}]{Doroshenko1995}
{Doroshenko} O.~V.,  {Kopeikin} S.~M.,  1995, \mn@doi [\mnras]
  {10.1093/mnras/274.4.1029}, \href
  {http://adsabs.harvard.edu/abs/1995MNRAS.274.1029D} {274, 1029}

\bibitem[\protect\citeauthoryear{{Edwards}, {Hobbs}  \& {Manchester}}{{Edwards}
  et~al.}{2006}]{Edwards2006}
{Edwards} R.~T.,  {Hobbs} G.~B.,   {Manchester} R.~N.,  2006, \mn@doi [\mnras]
  {10.1111/j.1365-2966.2006.10870.x}, \href
  {http://adsabs.harvard.edu/abs/2006MNRAS.372.1549E} {372, 1549}

\bibitem[\protect\citeauthoryear{{Falcke}, {Melia}  \& {Agol}}{{Falcke}
  et~al.}{2000}]{Falcke2000}
{Falcke} H.,  {Melia} F.,   {Agol} E.,  2000, \mn@doi [\apjl] {10.1086/312423},
  \href {http://adsabs.harvard.edu/abs/2000ApJ...528L..13F} {528, L13}

\bibitem[\protect\citeauthoryear{Fragione \& Loeb}{Fragione \&
  Loeb}{2020}]{Fragione2020}
Fragione G.,  Loeb A.,  2020, \mn@doi [The Astrophysical Journal]
  {10.3847/2041-8213/abb9b4}, 901, L32

\bibitem[\protect\citeauthoryear{{GRAVITY Collaboration} et~al.,}{{GRAVITY
  Collaboration} et~al.}{2020}]{Gravity2020}
{GRAVITY Collaboration} et~al., 2020, \mn@doi [A\&A]
  {10.1051/0004-6361/202037813}, 636, L5

\bibitem[\protect\citeauthoryear{Gates, Hadar  \& Lupsasca}{Gates
  et~al.}{2021}]{Gates2021}
Gates D. E.~A.,  Hadar S.,   Lupsasca A.,  2021, \mn@doi [Phys. Rev. D]
  {10.1103/PhysRevD.103.044050}, 103, 044050

\bibitem[\protect\citeauthoryear{Genzel, Sch{\"o}del, Ott, Eckart, Alexander,
  Lacombe, Rouan  \& Aschenbach}{Genzel et~al.}{2003}]{Genzel2003}
Genzel R.,  Sch{\"o}del R.,  Ott T.,  Eckart A.,  Alexander T.,  Lacombe F.,
  Rouan D.,   Aschenbach B.,  2003, \mn@doi [Nature] {10.1038/nature02065},
  425, 934

\bibitem[\protect\citeauthoryear{{Ghez} et~al.,}{{Ghez}
  et~al.}{2008}]{Ghez2008}
{Ghez} A.~M.,  et~al., 2008, \mn@doi [\apj] {10.1086/592738}, \href
  {http://adsabs.harvard.edu/abs/2008ApJ...689.1044G} {689, 1044}

\bibitem[\protect\citeauthoryear{{Gillessen}, {Eisenhauer}, {Trippe},
  {Alexander}, {Genzel}, {Martins}  \& {Ott}}{{Gillessen}
  et~al.}{2009}]{Gillessen2009}
{Gillessen} S.,  {Eisenhauer} F.,  {Trippe} S.,  {Alexander} T.,  {Genzel} R.,
  {Martins} F.,   {Ott} T.,  2009, \mn@doi [\apj]
  {10.1088/0004-637X/692/2/1075}, \href
  {http://adsabs.harvard.edu/abs/2009ApJ...692.1075G} {692, 1075}

\bibitem[\protect\citeauthoryear{{Goddi} et~al.,}{{Goddi}
  et~al.}{2017}]{Goddi2017}
{Goddi} C.,  et~al., 2017, \mn@doi [International Journal of Modern Physics D]
  {10.1142/S0218271817300014}, \href
  {http://adsabs.harvard.edu/abs/2017IJMPD..2630001G} {26, 1730001}

\bibitem[\protect\citeauthoryear{{Grould}, {Meliani}, {Vincent},
  {Grandcl{\'e}ment}  \& {Gourgoulhon}}{{Grould} et~al.}{2017}]{Grould2017}
{Grould} M.,  {Meliani} Z.,  {Vincent} F.~H.,  {Grandcl{\'e}ment} P.,
  {Gourgoulhon} E.,  2017, \mn@doi [Classical and Quantum Gravity]
  {10.1088/1361-6382/aa8d39}, \href
  {http://adsabs.harvard.edu/abs/2017CQGra..34u5007G} {34, 215007}

\bibitem[\protect\citeauthoryear{Hackmann}{Hackmann}{2010}]{Hackmann2010}
Hackmann E.,  2010, dissertation, Bremen university, \url
  {https://media.suub.uni-bremen.de/handle/elib/2804}

\bibitem[\protect\citeauthoryear{Hackmann \& Dhani}{Hackmann \&
  Dhani}{2019}]{Hackmann2019}
Hackmann E.,  Dhani A.,  2019, \mn@doi [General Relativity and Gravitation]
  {10.1007/s10714-019-2517-2}, 51, 37

\bibitem[\protect\citeauthoryear{{Hees} et~al.,}{{Hees}
  et~al.}{2017}]{Hees2017}
{Hees} A.,  et~al., 2017, \mn@doi [Physical Review Letters]
  {10.1103/PhysRevLett.118.211101}, \href
  {http://adsabs.harvard.edu/abs/2017PhRvL.118u1101H} {118, 211101}

\bibitem[\protect\citeauthoryear{Izmailov, Zhdanov, Bhadra  \&
  Kanti~Nandi}{Izmailov et~al.}{2019}]{Izmailov2019}
Izmailov R.,  Zhdanov E.,  Bhadra A.,   Kanti~Nandi K.,  2019, \mn@doi [The
  European Physical Journal C] {10.1140/epjc/s10052-019-6618-6}, 79

\bibitem[\protect\citeauthoryear{J.~M.~Bardeen}{J.~M.~Bardeen}{1972}]{Bardeen1972}
J.~M.~Bardeen W. H.~Press S. A.~T.,  1972, \mn@doi [Phys. Rev.]
  {10.1086/151796}, 178, 347

\bibitem[\protect\citeauthoryear{Keane et~al.}{Keane et~al.}{2015}]{Keane2014}
Keane E.~F.,  et~al., 2015, PoS, AASKA14, 040

\bibitem[\protect\citeauthoryear{{Kimpson, T.}, {Wu, K.}  \& {Zane,
  S.}}{{Kimpson, T.} et~al.}{2020}]{Kimpson2020}
{Kimpson, T.} {Wu, K.}  {Zane, S.} 2020, \mn@doi [A\&A]
  {10.1051/0004-6361/202038561}, 644, A167

\bibitem[\protect\citeauthoryear{Klioner}{Klioner}{1991}]{Klioner1991}
Klioner S.,  1991, Soviet Astronomy, 35, 523

\bibitem[\protect\citeauthoryear{Kopeikin}{Kopeikin}{1997}]{Kopeikin1997}
Kopeikin S.~M.,  1997, \mn@doi [Journal of Mathematical Physics]
  {10.1063/1.531997}, 38, 2587

\bibitem[\protect\citeauthoryear{Kramer}{Kramer}{2012}]{Kramer2012}
Kramer M.,  2012, in Neutron Stars and Pulsars: Challenges and Opportunities
  after 80 years. p.~19, \mn@doi{10.1017/S174392131202306X}, \url
  {http://journals.cambridge.org/article_S174392131202306X}

\bibitem[\protect\citeauthoryear{Lai \& Rafikov}{Lai \&
  Rafikov}{2005}]{Lai2005}
Lai D.,  Rafikov R.~R.,  2005, \mn@doi [Astrophys. J.] {10.1086/429146}, 621,
  L41

\bibitem[\protect\citeauthoryear{{Liu}, {Wex}, {Kramer}, {Cordes}  \&
  {Lazio}}{{Liu} et~al.}{2012}]{Liu2012}
{Liu} K.,  {Wex} N.,  {Kramer} M.,  {Cordes} J.~M.,   {Lazio} T.~J.~W.,  2012,
  \mn@doi [\apj] {10.1088/0004-637X/747/1/1}, \href
  {http://adsabs.harvard.edu/abs/2012ApJ...747....1L} {747, 1}

\bibitem[\protect\citeauthoryear{Liu, Eatough, Wex  \& Kramer}{Liu
  et~al.}{2014}]{Liu2014}
Liu K.,  Eatough R.~P.,  Wex N.,   Kramer M.,  2014, \mn@doi [Monthly Notices
  of the Royal Astronomical Society] {10.1093/mnras/stu1913}, 445, 3115

\bibitem[\protect\citeauthoryear{Lorimer \& Kramer}{Lorimer \&
  Kramer}{2005}]{Lorimer2005}
Lorimer D.~R.,  Kramer M.,  2005, Handbook of Pulsar Astronomy.
Cambridge University Press

\bibitem[\protect\citeauthoryear{{Lu}, {Broderick}, {Baron}, {Monnier}, {Fish},
  {Doeleman}  \& {Pankratius}}{{Lu} et~al.}{2014}]{Lu2014}
{Lu} R.-S.,  {Broderick} A.~E.,  {Baron} F.,  {Monnier} J.~D.,  {Fish} V.~L.,
  {Doeleman} S.~S.,   {Pankratius} V.,  2014, \mn@doi [\apj]
  {10.1088/0004-637X/788/2/120}, \href
  {http://adsabs.harvard.edu/abs/2014ApJ...788..120L} {788, 120}

\bibitem[\protect\citeauthoryear{{Mizuno} et~al.,}{{Mizuno}
  et~al.}{2018}]{Mizuno2018}
{Mizuno} Y.,  et~al., 2018, \mn@doi [Nature Astronomy]
  {10.1038/s41550-018-0449-5}, \href
  {http://adsabs.harvard.edu/abs/2018NatAs.tmp...41M} {}

\bibitem[\protect\citeauthoryear{NAN et~al.,}{NAN et~al.}{2011}]{Nan2011}
NAN R.,  et~al., 2011, \mn@doi [International Journal of Modern Physics D]
  {10.1142/S0218271811019335}, 20, 989

\bibitem[\protect\citeauthoryear{O'Neill}{O'Neill}{2014}]{O'Neill2014}
O'Neill B.,  2014, The geometry of Kerr black holes.
Courier Corporation

\bibitem[\protect\citeauthoryear{Pearlman, Majid  \& Prince}{Pearlman
  et~al.}{2019}]{Pearlman2019}
Pearlman A.~B.,  Majid W.~A.,   Prince T.~A.,  2019, \mn@doi [Advances in
  Astronomy] {10.1155/2019/6325183}, 2019, 6325183

\bibitem[\protect\citeauthoryear{{Psaltis}, {Wex}  \& {Kramer}}{{Psaltis}
  et~al.}{2016}]{Psaltis2016}
{Psaltis} D.,  {Wex} N.,   {Kramer} M.,  2016, \mn@doi [\apj]
  {10.3847/0004-637X/818/2/121}, \href
  {http://adsabs.harvard.edu/abs/2016ApJ...818..121P} {818, 121}

\bibitem[\protect\citeauthoryear{Rafikov \& Lai}{Rafikov \&
  Lai}{2005}]{Rafikov2005}
Rafikov R.,  Lai D.,  2005, \mn@doi [Physical Review D]
  {10.1103/PhysRevD.73.063003}, 73

\bibitem[\protect\citeauthoryear{{Rajwade}, {Lorimer}  \& {Anderson}}{{Rajwade}
  et~al.}{2017}]{Rajwade2017}
{Rajwade} K.~M.,  {Lorimer} D.~R.,   {Anderson} L.~D.,  2017, \mn@doi [\mnras]
  {10.1093/mnras/stx1661}, \href
  {http://adsabs.harvard.edu/abs/2017MNRAS.471..730R} {471, 730}

\bibitem[\protect\citeauthoryear{Schneider}{Schneider}{1990}]{Schneider1990}
Schneider J.,  1990, Astron. Astrophys., 232, 62

\bibitem[\protect\citeauthoryear{Sch{\"o}del et~al.,}{Sch{\"o}del
  et~al.}{2002}]{Schoedel2002}
Sch{\"o}del R.,  et~al., 2002, \mn@doi [Nature] {10.1038/nature01121}, 419, 694

\bibitem[\protect\citeauthoryear{{Semer{\'a}k}}{{Semer{\'a}k}}{2015}]{Semerak2015}
{Semer{\'a}k} O.,  2015, \mn@doi [\apj] {10.1088/0004-637X/800/1/77}, \href
  {http://adsabs.harvard.edu/abs/2015ApJ...800...77S} {800, 77}

\bibitem[\protect\citeauthoryear{{Torne, P.} et~al.,}{{Torne, P.}
  et~al.}{2021}]{Torne2021}
{Torne, P.} et~al., 2021, \mn@doi [A\&A] {10.1051/0004-6361/202140775}, 650,
  A95

\bibitem[\protect\citeauthoryear{{Verbiest} et~al.,}{{Verbiest}
  et~al.}{2008}]{Verbiest2008}
{Verbiest} J.~P.~W.,  et~al., 2008, \mn@doi [\apj] {10.1086/529576}, \href
  {http://adsabs.harvard.edu/abs/2008ApJ...679..675V} {679, 675}

\bibitem[\protect\citeauthoryear{Viergutz}{Viergutz}{1993}]{Viergutz1993}
Viergutz S.~U.,  1993, \mn@doi [Astronomy and Astrophysics]
  {1993A&A...272..355V}, 272, 355

\bibitem[\protect\citeauthoryear{{Vincent}, {Meliani}, {Grandcl{\'e}ment},
  {Gourgoulhon}  \& {Straub}}{{Vincent} et~al.}{2016}]{Vincent2016b}
{Vincent} F.~H.,  {Meliani} Z.,  {Grandcl{\'e}ment} P.,  {Gourgoulhon} E.,
  {Straub} O.,  2016, \mn@doi [Classical and Quantum Gravity]
  {10.1088/0264-9381/33/10/105015}, \href
  {http://adsabs.harvard.edu/abs/2016CQGra..33j5015V} {33, 105015}

\bibitem[\protect\citeauthoryear{Wang, Jenet, Creighton  \& Price}{Wang
  et~al.}{2009a}]{Wang2008}
Wang Y.,  Jenet F.~A.,  Creighton T.,   Price R.~H.,  2009a, \mn@doi
  [Astrophys. J.] {10.1088/0004-637X/697/1/237}, 697, 237

\bibitem[\protect\citeauthoryear{Wang, Creighton, Price  \& Jenet}{Wang
  et~al.}{2009b}]{Wang2009}
Wang Y.,  Creighton T.,  Price R.~H.,   Jenet F.~A.,  2009b, \mn@doi
  [Astrophys. J.] {10.1088/0004-637X/705/2/1252}, 705, 1252

\bibitem[\protect\citeauthoryear{{Wex} \& {Kopeikin}}{{Wex} \&
  {Kopeikin}}{1999}]{Wex1999}
{Wex} N.,  {Kopeikin} S.~M.,  1999, \mn@doi [\apj] {10.1086/306933}, \href
  {http://adsabs.harvard.edu/abs/1999ApJ...514..388W} {514, 388}

\bibitem[\protect\citeauthoryear{Wharton, Chatterjee, Cordes, Deneva  \&
  Lazio}{Wharton et~al.}{2012}]{Wharton2012}
Wharton R.~S.,  Chatterjee S.,  Cordes J.~M.,  Deneva J.~S.,   Lazio T. J.~W.,
  2012, \mn@doi [The Astrophysical Journal] {10.1088/0004-637x/753/2/108}, 753,
  108

\bibitem[\protect\citeauthoryear{Witzel et~al.,}{Witzel
  et~al.}{2018}]{Witzel2018}
Witzel G.,  et~al., 2018, \mn@doi [The Astrophysical Journal]
  {10.3847/1538-4357/aace62}, 863, 15

\bibitem[\protect\citeauthoryear{{Zhang} \& {Saha}}{{Zhang} \&
  {Saha}}{2017}]{Zhang2017}
{Zhang} F.,  {Saha} P.,  2017, \mn@doi [\apj] {10.3847/1538-4357/aa8f47}, \href
  {https://ui.adsabs.harvard.edu/\#abs/2017ApJ...849...33Z} {849, 33}

\bibitem[\protect\citeauthoryear{{Zschocke} \& {Klioner}}{{Zschocke} \&
  {Klioner}}{2009}]{Zschocke2009}
{Zschocke} S.,  {Klioner} S.~A.,  2009, preprint, \href
  {http://adsabs.harvard.edu/abs/2009arXiv0904.3704Z} {} (\mn@eprint {arXiv}
  {0904.3704})

\makeatother
\end{thebibliography}
\bsp	
\label{lastpage}
\end{document}